\documentclass[twocolumn,trackchanges]{aastex7}

\hypersetup{linkcolor=blue,citecolor=blue,filecolor=cyan,urlcolor=blue}

\usepackage{amsmath}

\begin{document}

\title{Classification of Radio Backgrounds at Cosmic Dawn and 21 cm Signal Confirmation Using Neural Networks}

\author[orcid=0000-0001-6129-0118,sname='Sikder']{Sudipta Sikder}
%\altaffiliation{Kitt Peak National Observatory}
\affiliation{School of Physics and Astronomy, Tel-Aviv University, Tel-Aviv, 69978, Israel}
\email[show]{sudiptas@mail.tau.ac.il} 

\author[orcid=0000-0002-1369-633X, sname='Fialkov']{Anastasia Fialkov} 
\affiliation{Institute of Astronomy, University of Cambridge, Madingley Road, Cambridge, CB3 0HA, UK}
\affiliation{Kavli Institute for Cosmology, Madingley Road, Cambridge, CB3 0HA, UK}
\email{afialkov@ast.cam.ac.uk}

\author[orcid=0000-0002-1557-693X,sname='Barkana']{Rennan Barkana}
\affiliation{School of Physics and Astronomy, Tel-Aviv University, Tel-Aviv, 69978, Israel}
\email{barkana@tauex.tau.ac.il} 

\begin{abstract}

Several ongoing and upcoming radio telescopes aim to detect either the global 21 cm signal or the 21 cm power spectrum. The extragalactic radio background, as detected by ARCADE-2 and LWA-1, suggests a strong radio background from cosmic dawn, which can significantly alter the cosmological 21 cm signal, enhancing both the global signal amplitude and the 21 cm power spectrum. In this paper, we employ an artificial neural network (ANN) to check if there is a radio excess over the cosmic microwave background in mock data, and if present, we classify its type into one of two categories, a background from high-redshift radio galaxies or a uniform exotic background from the early Universe. Based on clean data (without observational noise), the ANN can predict the type of background radiation with $96\%$ accuracy for the power spectrum and $90\%$ for the global signal. Although observational noise reduces the accuracy, the results remain quite useful. We also apply ANNs to map the relation between the 21 cm power spectrum and the global signal. By predicting  the global signal using the 21 cm power spectrum, an ANN can estimate the global signal range consistent with an observed power spectrum from SKA-like experiments. Conversely, we show that an ANN can predict the 21 cm power spectrum over a wide range of redshifts and wavenumbers given the global signal over the same redshifts. Such trained networks can potentially serve as valuable tools for cross-confirmation of the 21 cm signal.

\end{abstract}

\keywords{\uat{Early universe}{435}; \uat{Cosmology}{343}; \uat{HI line emission}{690}}


\section{Introduction} 

The 21 cm signal produced by neutral hydrogen atoms in the intergalactic medium is one of the most promising probes of the Epoch of Reionization (EoR) and cosmic dawn. Due to the expansion of the universe, this signal is redshifted to frequencies below 200 MHz and, therefore, is expected to be observed by low-frequency radio telescopes. The Experiment to Detect the Global EoR Signature (EDGES) reported the first tentative detection of the sky-averaged (global) 21 cm signal from $z \sim 17$ using the low band antenna in the $50-200$ MHz frequency range \citep{bowman18}. However, the SARAS3 experiment \citep{SARAS3} reported an upper limit of the global 21 cm signal which is inconsistent at $2\sigma$ with the EDGES signal. Other ongoing global experiments such as Sonda Cosmol´ogica de las Islas para la Detecci´on de Hidr´ogeno Neutrothe \citep[SCI-HI,][]{scihi2014}, Probing Radio Intensity at high-Z \citep[PRIZM,][]{philip19}, Mapper of the IGM Spin Temperature \citep[MIST,][]{mist2024}, Radio Experiment for the Analysis of Cosmic Hydrogen \citep[REACH,][]{de_lora2022} may soon validate or refute the purported EDGES claims. On another front, several observational efforts have been made to try to measure the spatial fluctuations in the 21 cm signal, i.e., the 21 cm power spectrum from the EoR and cosmic dawn. These used radio interferometers including the Low Frequency Array \citep[LOFAR,][]{lofar_gehlot},  the Murchison Wide-field Array \citep[MWA,][]{mwa_trott},  the Owens Valley Radio Observatory Long Wavelength Array \citep[OVRO-LWA,][]{lwa_eastwood}, the Large-aperture Experiment to detect the Dark Age \citep[LEDA,][]{leda_price, leda_garsden}, the New Extention in Nan\c{c}ay Upgrading LOFAR  \citep[NenuFAR,][]{mertens2021exploring}, and the Hydrogen Epoch of Reionization Array \citep[HERA,][]{hera_DeBoer, hera2022_first}. The upcoming Square Kilometre Array \citep[SKA,][]{Koopmans} is expected to provide measurements of the 21 cm power spectrum over a wide range of redshifts including cosmic dawn. 

Although semi-numerical simulations \citep[e.g.][]{mesinger11, visbal12, fialkov14b, cohen17} are significantly faster than full cosmological hydrodynamical simulations, each run can still take hours to calculate the 21 cm signal from a wide range of redshifts and large cosmological volume. To accurately constrain astrophysical parameters, using semi-numerical simulations directly is computationally demanding. Instead, emulators trained using an Artificial Neural Network (ANN) can be utilized to generate the desired statistical description of the signal in a fraction of a second. Such emulation methods drastically speed up the parameter explorations of astrophysical processes using observational data. Various works have been done on the emulation of the global 21 cm signal and the 21 cm power spectrum. The first ANN global signal emulator was introduced by \citet{cohen20}. Recently, different approaches based on ANN have been proposed to emulate the global signal more accurately \citep{globalemu,21cmvae}. \citet{Kern} proposed a fast emulator for the 21 cm power spectrum and performed Bayesian analysis to constrain an eleven parameter model that included six astrophysical parameters of reionization and X-ray heating and five additional cosmological parameters. \citet{Schmit} used an ANN to emulate the 21 cm power spectrum and compared their analysis to constraining three EoR parameters with 21CMMC \citep{Greig2015}. Some of these emulators have been used to find constraints on astrophysical parameters \citep{Mondal2020LOFAR, hera2022, hera2022_, bevins2022, Bevins2023}. In our previous work \citep{sikder}, we constructed an ANN-based emulation method to constrain uncertainties in a seven-parameter astrophysical model using a mock SKA data set of the 21 cm power spectrum, where we added the expected SKA noise and other observational effects to the simulated power spectrum. In addition, we considered astrophysical models with two types of excess radio backgrounds: external radio background models, which refer to a homogeneous radio background with a redshift-independent amplitude that is not directly associated with astrophysical sources, as proposed by \citet{fialkov19}; and galactic radio models, which refer to excess radio emission from high redshift galaxies, as suggested by \citet{Reis2020}. We built an ANN to infer the type of radio background present in the 21 cm power spectrum. However, we did not compare with models with a standard astrophysical scenario (where the Cosmic Microwave Background$-$CMB$-$is the only source of background radio emission) in the classification procedure. In this work, we construct an ANN classifier and apply the trained classifier on a data set of three different radio background models in order to identify the type of radio background in the 21 cm signal from cosmic dawn and the EoR. We show for the first time that the global 21 cm signal as well as the power spectrum over a wide range of redshifts can be used to infer the presence and type of radio background present in the signal using an ANN based classifier. 

Although fast progress in observational 21 cm cosmology is being made, we do not yet know whether the global signal or the power spectrum will be measured more precisely in the near future. So in the current early state of observations, an interesting theoretical question arises as to how well we can determine the global 21 cm signal given a measurement of the power spectrum from observations like the SKA. We address this question in the second part of this work. We present an ANN to predict the global 21 cm signal from cosmic dawn and the EoR given the 21 cm power spectrum over the same redshifts. This method could, in principle, test whether or not independent observations by radiometers and interferometers are consistent. We also consider the inverse direction. With the type of radio background successfully established from a global signal measurement using our classification method described in the previous paragraph, we show that another ANN can be used to predict the power spectrum. Using this trained network, we could potentially put constraints on the 21 cm power spectrum from an observed global signal. Such methods could be used to test the consistency between future detections of the global signal and power spectrum, and thus are useful for the purpose of verification of the 21 cm signal. 

This paper is organized as follows: In Section~\ref{sec:theory}, we describe the theoretical background of the astrophysics of the high redshift Universe and the 21 cm signal. We discuss the method to generate the data sets and the details of the neural network architecture in Section~\ref{sec:methods}. Sections \ref{sec:classification} and \ref{sec:mapping_21cm_observables} present the results of this work. Finally, we conclude in Section~\ref{sec:conclusion}.

\section{Theoretical Background}\label{sec:theory}
\subsection{Astrophysics at high redshift}\label{sec:parameters}
We use our simulation code 21 cm Seminumerical Predictions Across Cosmological Epochs \citep[21cmSPACE, e.g.][]{visbal12, fialkov14b, fialkov19, cohen17}. We simulate realizations of the universe in a $(384\,{\rm Mpc})^3$ comoving cosmological volume over the redshift range $6-35$ thus including cosmic dawn and reionization. Following hierarchical structure formation, the simulation keeps track of star formation, evolution of the X-ray and Ly$\alpha$ backgrounds, and other radiative feedback mechanisms (such as the effect of Lyman-Warner feedback and photoheating feedback on star formation). We parameterize the high redshift universe using eight key parameters which regulate several astrophysical processes including star formation, heating, ionization, and an excess radio background. 

Given the power spectra of initial density and velocity fields \citep[{calculated using the publicly available code \sc CAMB},][]{camb}, a random realization of the linear density field and relative velocity field between dark matter and baryons \citep{tseliakhovich10} are calculated on a $128^3$ grid with a resolution (pixel size) of $3$ comoving Mpc. The density and velocity fields are then evolved linearly. We obtain the population of collapsed dark matter halos in each cell at each redshift using a modified Press-Schechter model \citep{press74, sheth99, barkana04}. The dark matter halo population that can form stars is determined by the minimum circular velocity ($V_{\rm{c}}$) of the halos, a free parameter in the simulation. Another free parameter encoding the information about star formation is the fractional amount of gas in star-forming halos that is converted into stars, referred to as the star formation efficiency ($f_{\star}$). The simulation takes into account the suppression of star formation due to the above-mentioned relative velocity between dark matter and baryons, Lyman-Warner feedback \citep{haiman97,fialkov2013}, photoheating feedback during the EoR \citep{rees86, sobacchi13,cohen16}, and also Ly$\alpha$ and CMB heating \citep{reis2021}.  

Given a galaxy population, the simulation calculates radiation fields emitted by those galaxies. The intensity of the Ly$\alpha$ radiation is calculated assuming that galaxies contain population II stars. The dominant contribution to the X-ray luminosity ($L_{\rm{X}}$) in galaxies is likely to be from X-ray binaries. To parameterize the total X-ray luminosity of various sources, galactic halos are assumed to have an X-ray luminosity that scales with the star formation rate \citep[e.g.,][]{Grimm, Gilfanov, Mineo:2012, fragos13, fialkov14a, Pacucci:2014}. The scaling is such that the standard normalization factor (which we denote as an X-ray radiation efficiency $f_{\rm{X}} = 1$) corresponds to the X-ray luminosity of present day low-metallicity galaxies. As $f_{\rm{X}}$ has relatively weak observational constraints at the relevant redshifts \citep{fialkov2017, hera2022, Bevins2023}, we vary $f_{\rm{X}}$ from $0.0001$ to $1000$. In addition to the X-ray efficiency $f_{\rm{X}}$, the SED of the X-ray sources is another key astrophysical parameter. We assume that the X-ray SED has a power law shape with a slope $\alpha$ and low energy cutoff $E_{\rm{min}}$. We vary $\alpha$ in the range $1-1.5$ and $E_{\rm{min}}$ in the range $0.1-3.0$ keV. 

The process of reionization is parameterized with two parameters: one is the ionizing efficiency $\zeta$ of sources. Given that, for a specific choice of other astrophysical parameters, there exists a one-to-one relationship between the CMB optical depth ($\tau$) and the ionizing efficiency ($\zeta$), and as $\tau$ is directly probed by CMB observations \citep{planckcollaboration18}, we opt to incorporate $\tau$ as part of our astrophysical model instead of the ionizing efficiency ($\zeta$). Another EoR parameter is the mean free path of the ionizing photons, $R_{\rm{mfp}}$, which we vary between 10 and 70 comoving Mpc \citep{Wyithe_loeb2004,Songaila2010}. This parameter quantifies the maximum possible radius of an ionized bubble during the EoR, and is important only in the latter stages. For convenience, we list the astrophysical parameters (including those in the excess radio models) along with their allowed regions in Table. \ref{tab:parameters_list}. 

The above describes our standard astrophysics model. The simulation also includes the possibility of an excess radio background over the CMB. This feature is discussed in detail in Section~\ref{sec:excess_radio_models}.

\subsection{21 cm signal}

The 21 cm brightness temperature $T_{\rm{21}}$ measures the 21 cm intensity on the sky. In the low optical depth regime (where the 21 cm optical depth, $\tau_{\rm{21}} \ll 1$), the 21 cm brightness temperature can be approximated as \citep{Madau1997, Pritchard2012, barkana18book} 
\begin{multline}
\label{eqn:tb}
T_{\rm 21}  \approx  26.8 \left(\frac{\Omega_{\rm b} h}{0.0327}\right)
\left(\frac{\Omega_{\rm m}}{0.307}\right)^{-1/2}
\left(\frac{1 + z}{10}\right)^{1/2} \\
(1 + \delta) x_{\rm HI} \frac{x_{\rm tot}}{1 + x_{\rm tot}} \left( 1 - \frac{T_{\rm rad}}{T_{\rm K}}\right) ~{\rm mK}\ ,
\end{multline}
where $\delta$ is the density contrast, $x_{\rm{HI}}$ is the neutral hydrogen fraction, and the coupling coefficient $x_{\rm{tot}}$ is the sum of the contributions of the Ly$\alpha$ coupling ($x_{\alpha}$) and the collisional coupling ($x_{c}$), i.e., $x_{\rm{tot}} = x_{\alpha} + x_{c}$. The standard astrophysical scenario assumes the background radiation to be the CMB, in which case $T_{\rm{rad}} = T_{\rm{CMB}} = 2.725(1 + z)$ K.
In the presence of an excess radio background of brightness temperature $T_{\rm{Radio}}$, the total radio background can be written as 
\begin{equation}
T_{\rm{rad}} = T_{\rm{CMB}} + T_{\rm{Radio}}\ .
\end{equation}

There are two approaches to measuring the 21 cm signal. The first is to integrate over a large area on the sky with a single antenna and find the sky-averaged (global) spectrum, which tracks the evolution of the signal with time/redshift. The second approach is to use interferometry and measure the evolution of the fluctuations in the 21 cm signal. This yields a much richer data set that gives spatial information over a wide range of wavenumbers at each redshift. The 21 cm power spectrum of the brightness temperature fluctuations is defined as
\begin{equation}
\langle \tilde{\delta}_{T_b}(\mathbf{k})\tilde{\delta}^*_{T_b}(\mathbf{k^{\prime}}) \rangle = (2\pi)^3\delta_D(\mathbf{k}-\mathbf{k^{\prime}})P(\mathbf{k})\ ,
\end{equation}
where $\mathbf{k}$ is the comoving wave vector, $\delta _D$ is the Dirac delta function, $\tilde{\delta}_{T_b} (\mathbf{k})$ is the Fourier transform of $\delta_{T_b} (\mathbf{x})$ which is defined by $\delta_{T_b} (\mathbf{x}) = ( T_{21}(\mathbf{x}) - \langle T_{21} \rangle)/\langle T_{21} \rangle$, and the angular brackets denote the ensemble or spatial average. It is conventional to express the power spectrum in terms of the variance in mK$^2$ units as follows:
\begin{equation}
\Delta^2 = \langle T_{21} \rangle ^2 \frac{k^3P(k)}{2\pi^2}\ .
\end{equation}

Although the 21 cm power spectrum does not capture all the available statistical information, due to the non-Gaussian fluctuations sourced by non-linear processes over both large and small scales during cosmic dawn and the EoR, it remains an invaluable tool to reveal significant astrophysical insights. Its measurement is relatively straightforward through observations compared to higher order statistics such as the 21 cm bispectrum \citep{Lewis2011, Majumdar2018}, trispectrum \citep{Cooray2008}, etc.

\subsection{Excess radio background models}\label{sec:excess_radio_models}
As mentioned in the introduction, though not yet independently confirmed, a possible detection of the 21 cm signal from cosmic dawn ($z \sim 17$) has been claimed by the EDGES collaboration \citep{bowman18}. The reported signal has an absorption trough which is a factor of two larger in amplitude than what is predicted by a standard astrophysical model based on the $\Lambda$CDM cosmology and hierarchical structure formation \citep{cohen17, reis2021}. This unexpectedly strong absorption signature, if it is cosmological (rather than instrumental or due to foregrounds), signifies a large difference between the spin temperature of neutral hydrogen and the temperature of the radiation background. A  modification of the standard astrophysical model is required to explain the anomalous absorption signal. \citet{barkana18} showed that interaction of cold dark matter with the cosmic gas could introduce an additional cooling mechanism, making the neutral hydrogen gas colder than expected \citep[see also][]{Berlin, munoz18, Liu19, Cheung2019, Barkana2018_, barkana2022, Kovetz2018}.
Another potential explanation involves the presence of an excess radio background over the CMB \citep{bowman18, feng18, ewall18, fialkov19, mirocha19, ewall20, Reis2020}. In this paper we focus on the second explanation and explore two different types of previously-developed excess radio background models. The first is an assumed homogeneous background that was produced by exotic processes in the early universe and then passively redshifted in the Universe during observable epochs, which we will call the ``external radio background." The second is a background produced by the same galaxies that are responsible for all other radiation backgrounds (except the CMB); in this case, the background is naturally inhomogeneous and dynamically evolving, and we will refer to it as the ``galactic radio background." We also note that recent studies \citep{acharya23, cyr24} indicate that soft photon heating may alter the shape and amplitude of the 21 cm signal in the presence of a strong radio background during cosmic dawn. This effect is not accounted for in this work, but we intend to include it in future work. This is a point to which we return in Section~\ref{sec:mapping_21cm_observables}.

\ \ $\bullet$ \textbf{External radio background:} \citet{fialkov19} proposed that the EDGES detection, if confirmed, can be explained by an excess homogeneous radio background with a redshift-independent amplitude and synchrotron-like spectrum in the early Universe \citep[see ][for a recent review]{Singal2023PASP}. This background would not be directly related to astrophysical sources and could instead be generated by exotic processes such as dark matter decay. The total radio background due to this phenomenological uniform excess, with a form that is inspired by ARCADE2 \citep{fixsen11, seiffert11} and LWA1 \citep{dowell18} observations, can be written as  
\begin{equation}\label{eq:Texternal}
T_{\rm{rad}} =  T_{\rm{CMB}}(1 + z)\left[ 1 + A_{\rm{r}} \left( \frac{\nu_{\rm{obs}}}{78 \ \rm{MHz}}\right)^{\beta}\right] \ \hbox{K} \ ,
\end{equation}
where $ T_{\rm{CMB}}$ is the CMB temperature today, $\nu_{\rm{obs}}$ is the observed frequency, the spectral index $\beta = -2.6$ and the constant $A_{\rm{r}}$ is the amplitude of this uniform radio background relative to the CMB. Thus, in total, we have eight free parameters for the external radio model: $f_{\star}$, $V_{\rm{c}}$, $f_{\rm{X}}$, $\alpha$, $E_{\rm{min}}$, $\tau$, $R_{\rm{mfp}}$ and $A_{\rm{r}}$. 

\ \ $\bullet$ \textbf{Galactic radio background:} In contrast with the phenomenological external radio model, a fluctuating radio excess could originate from astrophysical sources such as active galactic nuclei  \citep[AGN,][]{urry95, biermann14, bolgar18, ewall18, ewall20} or star-forming galaxies \citep{condon92, jana19} at high redshift. The effect of this inhomogeneous excess galactic radio background on the global 21 cm signal and the 21 cm power spectrum was explored by \citet{Reis2020}, who first incorporated it into seminumerical simulations of the early Universe and whose models we use here\footnote{Note that in our more recent work \citep{sikder2023}, we presented an improved model of the nonuniform radio fluctuations incorporating a line-of-sight effect; this effect is not included in the current analysis.}. The galactic radio background is calculated assuming the galaxy radio luminosity per unit frequency to be proportional to the star formation rate (SFR):
\begin{multline}
L_{\rm{Radio}}(\nu, z) = f_{\rm{Radio}} \times 10^{22} \left(\frac{\nu}{150~\hbox{MHz}} \right)^{-\alpha_{\rm{Radio}}} \\
\left(\frac{\rm{SFR}}{\rm{M_\odot yr^{-1}}}\right) \ \ \ \ \rm{W\,Hz}^{-1}\ ,
\end{multline}
where SFR is the star formation rate, the spectral index in the radio band $\alpha_{\rm{Radio}}$ is set to the typical value of $0.7$ \citep{mirocha19, gurkan18} and $f_{\rm{Radio}}$ is the normalization of the radio emissivity. A value of $f_{\rm{Radio}} = 1$ corresponds to present day star-forming galaxies. Thus, the eight free parameters for the galactic radio models are: $f_{\star}$, $V_{\rm{C}}$, $f_{\rm{X}}$, $\alpha$, $E_{\rm{min}}$, $\tau$, $R_{\rm{mfp}}$, and $f_{\rm{Radio}}$.

\begin{table*}
\centering
\begin{tabular}{lcc} 
\hline
Astrophysical parameters    & Allowed range  & Description \\ 
\hline\hline
$f_{\star}$  & $0.001 - 0.5$     & Star formation efficiency \\
$V_{\rm{c}}$  & $4.2 - 100$ km s$^{-1}$    & Minimum circular velocity \\
$f_{\rm{X}}$  & $10^{-4} - 10^3$     & X-ray production efficiency \\
$\alpha$  & $1-1.5$      & Slope of X-ray SED  \\
$E_{\rm{min}}$  & $0.1-3.0$ keV   & X-ray SED low energy cutoff\\
$\tau$  & $0.028-0.098$     & CMB optical depth \\
$R_{\rm{mfp}}$  & $10-70$ Mpc  & Mean free path for ionizing photons \\
$f_{\rm{Radio}}$  & $0.01-10^6$    & Radio production efficiency (Galactic radio model)\\
$A_{\rm{r}}$  & $10^{-4}-0.5$    & Amplitude of uniform radio background (External radio model)\\

\hline
\end{tabular}
\caption{The astrophysical parameters and their allowed ranges in our full data set that includes standard astrophysical (CMB only), external and galactic radio models.}\label{tab:parameters_list}

\end{table*}

Although both an external and a galactic radio excess has an impact on the 21 cm signal, resulting in a significant enhancement of the signal during cosmic dawn and the epoch of reionization, there are notable distinctions between the two models. In our assumed external radio background model, it is spatially uniform, is present during early cosmic epochs even prior to cosmic dawn (along with the CMB), and its strength decreases with cosmic time. In contrast, the excess radio background from high redshift galaxies is inhomogeneous and its intensity increases with cosmic time as it traces the formation and growth of galaxies, assuming that $f_{\rm{Radio}}$ remains constant with redshift. We expect these differences to give an ANN a chance to distinguish signatures of these radio backgrounds in the simulated 21 cm signals.

\section{Methods}\label{sec:methods}

\subsection{Methods to generate the data sets}

%\textcolor{red}{We used uniform random sampling in linear space for all parameters unless otherwise noted (e.g., [specific parameter] sampled log-uniformly due to its wide range).}

In this paper, we utilize three of our archival data sets of 21 cm signals created using 21cmSPACE for a large number of astrophysical parameter combinations and across a wide range of redshifts. The parameters are sampled randomly across wide prior ranges. The first data set was created using the CMB as the sole radio background, i.e., no additional radio background was present (we refer to this case as the {\it standard astrophysical scenario}). Additionally, we include two data sets of models that incorporate an excess radio background, either galactic or extragalactic. The standard astrophysical data set includes 3193 models (21 cm power spectra and corresponding global signals) covering a broad range of the seven astrophysical parameters, as detailed in Section~\ref{sec:parameters}. In the case of the excess radio background cases, where the number of free parameters increases from seven to eight, the data sets consist of 10032 and 5075 models for the galactic radio background and the external radio background, respectively. However, to avoid any classification biases during the training process, we use a similar number of models (approximately 3,000 randomly selected models from each category) for the classification analysis. Although the number of models in these archival data sets is modest, it is enough for the proof-of-concept analysis which we report in this paper. To simulate a more realistic observational scenario, we use one more data set of 21 cm power spectra with expected observational effects from an SKA-like experiment, referred to as `mock SKA data' (as explained in Section~\ref{sec:noise_data}). Our full simulated redshift range (both for the global signals and the power spectra) is $z=6 - 35$, with 30 redshift bins ($\Delta z = 1$). For the theoretical power spectrum (without observational effects) we use 32 $k$-bins that cover the range 0.0492 to 1.095~Mpc$^{-1}$. Throughout this work, we adopt the $\Lambda$CDM cosmology with cosmological parameters from \citet{planck2014}.

We note that the size of our simulated sample is relatively small compared to the dimensionality of the full astrophysical parameter space. Nevertheless, we argue that the current sample size is sufficient for training purposes, as it spans a representative range of outputs that enables the model to generalize effectively, as evidenced by the performance metrics reported in Section \ref{sec: cross-validation}. The objective of our sampling strategy was to ensure representative coverage of the relevant astrophysical parameter space, rather than achieving exhaustive coverage of the full astrophysical parameter space. While random sampling inherently lacks the uniformity of structured approaches, it provides adequate diversity in the training data to capture the key variations in model outputs. The robustness of the predictions of our trained model suggests that adopting a more structured sampling method would not substantially alter the results. We leave for further future work a systematic investigation of the impact of different sampling strategies—such as grid-based sampling, random sampling, and Latin hypercube sampling—on the performance of the ANN classification models, as well as on the predictive capability between observables.

\subsection{Mock observational data sets}\label{sec:noise_data}

In order to consider more realistic cases, we include several expected observational effects in the 21 cm power spectrum. In order to generate the mock SKA data set, we process the direct output of the simulation using the following prescription: (a) Two-dimensional slices of simulation boxes (images) are smoothed with a two-dimensional Gaussian that corresponds to the effect of SKA resolution. (b) We add a pure Gaussian noise (smoothed with the SKA resolution) to the image as a realization of the SKA thermal noise. (c) We adopt a mild foreground avoidance to mitigate the foreground effects following \citet{discrete} \citep[see also ][]{datta10, dillon14, pober14, pober15, jensen15}. The expected redshift dependence of these three effects is included. Foreground avoidance refers to the assumption that a region of the observed $k$-space must be dropped from the analysis since it is dominated by foregrounds; by ``mild" we mean that this avoidance region is a relatively small wedge-like region, assuming that the SKA will enable a reasonably accurate foreground removal as a first step of foreground mitigation. Specifically, we assume that the wedge is determined by the SKA field of view, corresponding to the ``optimistic model" from \citet{pober14}. For the `mock SKA data', we utilize eight redshift bins (covering $z=6 - 27.4$) and five $k$ bins (covering $k=0.05$~Mpc$^{-1}$ to 1~Mpc$^{-1}$). At each redshift, we calculate  the mean of the 21 cm power spectrum over the range of $k$ values within each $k$-bin. This procedure is applied to all models in the mock SKA data set \citep[for more details, see][]{sikder}. We do not include detailed interferometer effects that are related to calibration and limited Fourier coverage.

To create a realistic data set for the global 21 cm signal, we introduce random Gaussian noise to the ideal global signal generated by the simulation. Specifically, to the signal in each redshift bin ($\Delta z = 1$) we add Gaussian noise with mean $\mu = 0$ and $\sigma$ of 2.5 mK, 17 mK, or 25 mK, resulting in three levels of noisy global 21 cm signal data sets. The 17 mK and 25 mK noise levels correspond to the range of typical sensitivity of existing telescopes \citep{bowman18, de_lora2022}, while the 2.5 mK level is an optimistic scenario that represents the sensitivity of a next-generation global signal telescope. Below we sometimes also restrict the considered redshift range when considering various specific experiments.

\subsection{ANN architectures,  hyper-parameter tuning, and data pre-processing }

We employ a feed-forward artificial neural network (ANN) in the form of a Multi-layer Perceptron \citep[MLP;][]{Ramchoun2016MultilayerPA}, implemented in the \texttt{Scikit-learn} library \citep{scikit_learn}. The network consists of an input layer, one or more hidden layers with non-linear activation functions, and an output layer. The MLP is trained using backpropagation \citep{back_propagation} with gradient-based optimization, minimizing the mean squared error for regression tasks or cross-entropy loss for classification tasks. Training starts from randomly initialized weights and continues until convergence or until a maximum number of iterations is reached.

The number of input neurons depends on the specific data set used. For classification using the theoretical 21 cm power spectrum without observational effects, the inputs are the power spectra over 32 $k$-bins ($k=0.0492–1.095$ Mpc$^{-1}$) for each of the 30 redshift bins ($z=6–35$, $\Delta z=1$), flattened into a single vector. For radio background classification with the `mock SKA' 21 cm power spectrum, the input layer has 40 neurons corresponding to the 8 redshift bins ($z=6–27.4$) and 5 $k$-bins ($k=0.05–1.0$ Mpc$^{-1}$), also flattened into a single vector. For classification using the global 21 cm signal, the inputs are the 30 redshift bins covering $z=6–35$ ($\Delta z=1$).

Regardless of the input configuration, for our three-class classification task, distinguishing between no excess radio background, an external radio background, and a galactic radio background, the output layer contains three neurons, one corresponding to each class. A softmax activation function is applied to these outputs to yield class probabilities, and the predicted class is taken to be the one with the highest probability.

Although an MLP has the capability to learn highly non-linear models, it is sensitive to feature scaling and requires tuning several hyper-parameters. In an ANN, hyper-parameters are those that cannot be optimized during the training process. These include the number of hidden layers, the number of nodes/neurons in each layer, the activation function for hidden layers, and the solver for weight optimization. To perform hyper-parameter tuning, we use \textit{GridSearchCV}, which exhaustively searches all hyper-parameter combinations over a specified grid, fits the model, and retains the best combination of hyper-parameters.  

Data pre-processing is a crucial step in enhancing the performance of an ANN. We log-transform the power spectrum values to linearize the data and apply either \textit{StandardScaler} or \textit{QuantileTransformer} from the \texttt{Scikit-learn} library \citep{scikit_learn} to scale the data to have a mean of 0 and a standard deviation of 1. For each category of data (global signal, clean power spectrum, and mock SKA power spectrum), standardization is performed once on all models in the training data set, including models from various classes. Standardizing the combined data set preserves information on the relative amplitudes between different model classes. In each case, we select the utility function that yields the highest classification accuracy.

Although we use the full 21 cm power spectrum across all $k$- and $z$-bins for classification of radio backgrounds, we reduce the dimensionality of these data sets when predicting the power spectrum from the global 21 cm signal, or vice versa. When dealing with hundreds of data outputs, such as power spectra measured in multiple $k$-modes across numerous redshifts, creating a separate predictive model or emulator for each output becomes computationally intensive and complex. One approach to mitigate this is to resample the data onto a coarser grid using interpolation techniques, such as linear interpolation or cubic splines. This reduces the number of inputs to the ANN, decreasing both the size of the network and the computational complexity. Alternatively, Principal Component Analysis (PCA), a fast and unsupervised algorithm \citep{pca}, offers an effective solution by compressing the dimensionality of the data while preserving its most critical information. In the 21 cm power spectra, the neighboring $k$- and $z$-bins are highly correlated due to the smoothness of the spectra, indicating redundancy across these dimensions. PCA exploits this correlation by transforming the data into a set of independent principal components (PCs), linear combinations of the original variables that capture the maximum variance, so that the original data can be approximated with fewer variables. Rather than predicting every element of the data vector (e.g., each $k$-mode at every redshift), we emulate the weights of these PCs. PCA thus compresses the data by projecting it into a lower-dimensional space of independent components. Previous studies \citep{Kern, sikder} have successfully applied PCA to preprocess power spectrum data sets for machine learning algorithms. Inspired by this, we employ PCA as a third step in our data preprocessing pipeline. However, selecting too many PCs can lead to overfitting, where the model becomes overly tailored to the training data and fails to generalize to new data. To avoid this, we determine the optimal number of PCs based on two regression performance metrics: the $R^2$ score and mean squared error (MSE), using functions from the \texttt{Scikit-learn} metrics module. Details on the number of PCA components used for signal reconstructions (either the power spectrum or global signal) are provided later (in Sections \ref{sec:gs_prediction} and \ref{sec:ps_prediction}). In Table \ref{tab:table_ANN}, we present input features, hidden layers, activation function and outputs for all classification and prediction networks considered in this work. The specific network architectures and preprocessing steps for each network are discussed in the following section.

\begin{table*}
\centering
\begin{tabular}{lcccc} 
\hline
ANN classification  & Input features & Hidden layers & Activation function & Output \\ 
\hline\hline
 & Ideal PS in 30 $z$ \& 32 $k$-bins  & (134, 134, 134) & ReLU & Predicted class/type  \\
 & Ideal GS in 30 $z$-bins  & (143) & Logistic sigmoid & Same \\
\hline
 & `Mock SKA' PS in 8 $z$ \& 5 $k$-bins & (200, 200, 200) & ReLU & Same \\
 & GS with Gaussian noise  & (143) & Logistic sigmoid & Same \\

\hline
\hline
ANN Prediction  & Input features & Hidden layers & Activation function  & Output \\ 
\hline\hline
 & 100 PCs of Ideal PS  & (150, 150, 150) & ReLU & Ideal GS in 30 $z$-bins \\
 & `Mock SKA' PS in 8 $z$ \& 5 $k$-bins & (150, 150, 150) & ReLU & Same \\
\hline
 & Ideal GS in 30 $z$-bins  & (300, 300, 300, 300) & ReLU & 22 PCs of Ideal PS  \\
 & GS with Gaussian noise ($\sigma=2.5$)  & (300, 300, 300, 300) & ReLU & 50 PCs of Ideal PS \\
 & GS with Gaussian noise ($\sigma=17$)  & (300, 300, 300, 300) & ReLU & 215 PCs of Ideal PS \\

\hline
\hline

\end{tabular}
\caption{Input features, hidden layers, activation functions and outputs for all ANNs considered in the work. Here PS $=$ power spectrum and GS $=$ global signal.}
\label{tab:table_ANN}
\end{table*}

\begin{figure}
    \centering
    \includegraphics[scale=0.5]{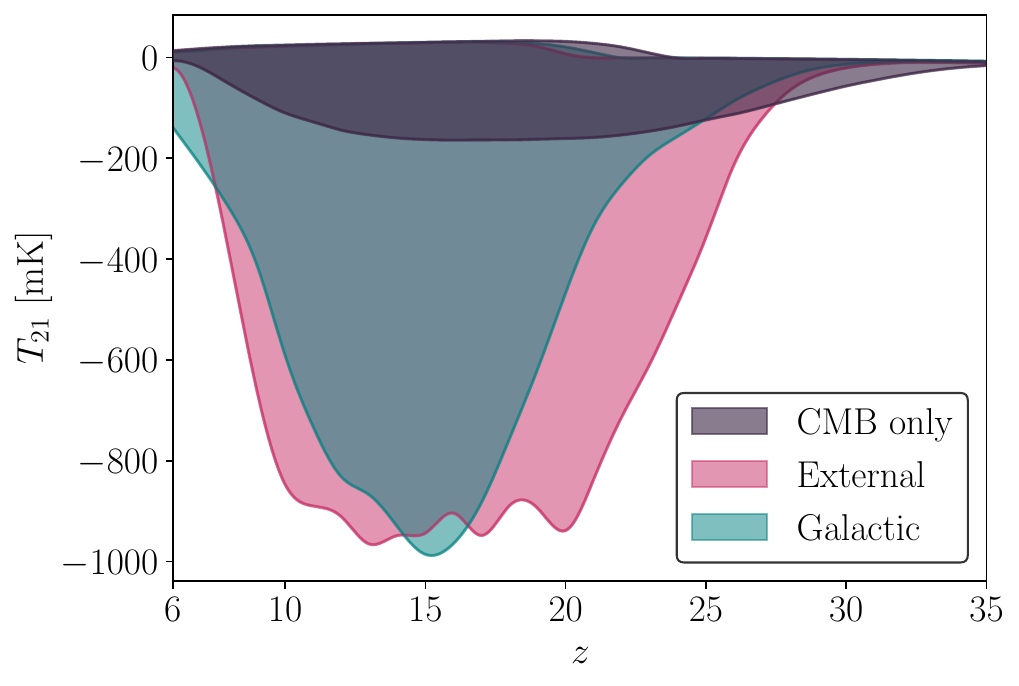}
    \includegraphics[scale=0.5]{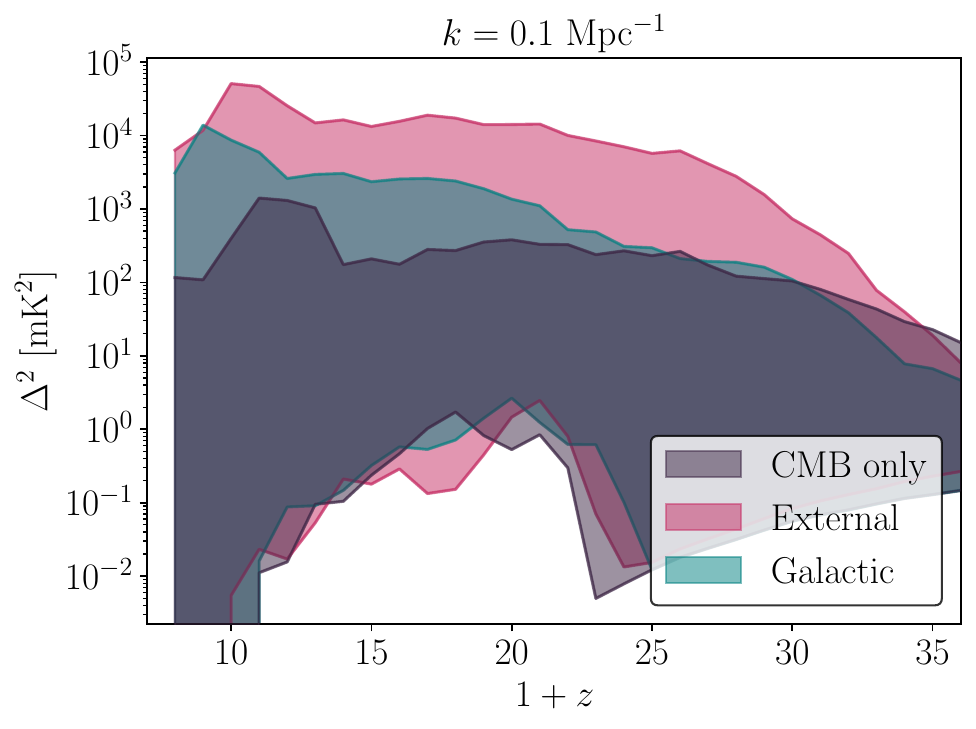}
    \caption{\textbf{Top panel:} The global signal envelopes for three different classes of models included in our test data set for classification. \textbf{Bottom panel:} The envelopes for the 21 cm power spectrum at $k = 0.1$ Mpc$^{-1}$. Here we show the theoretical 21 cm signal, without any observational effects.}
    %\caption{The global signals (top panel) and the 21-cm power spectra (bottom panel) at $k = 0.1$ Mpc$^{-1}$ for the three different classes of models included in our test dataset for classification. Here we show the theoretical 21-cm signal, without any observational effects.}
    \label{fig:test_models}
\end{figure}

\section{Classification of the radio background}\label{sec:classification}

\subsection{Classification using the ideal data set}
\label{sec:classification_clean}
\begin{figure*}
    \centering
    \begin{minipage}{0.4\textwidth}
    \centering
    \hspace{1.5cm}
    \textbf{Power spectrum}
       \includegraphics[scale=0.5]{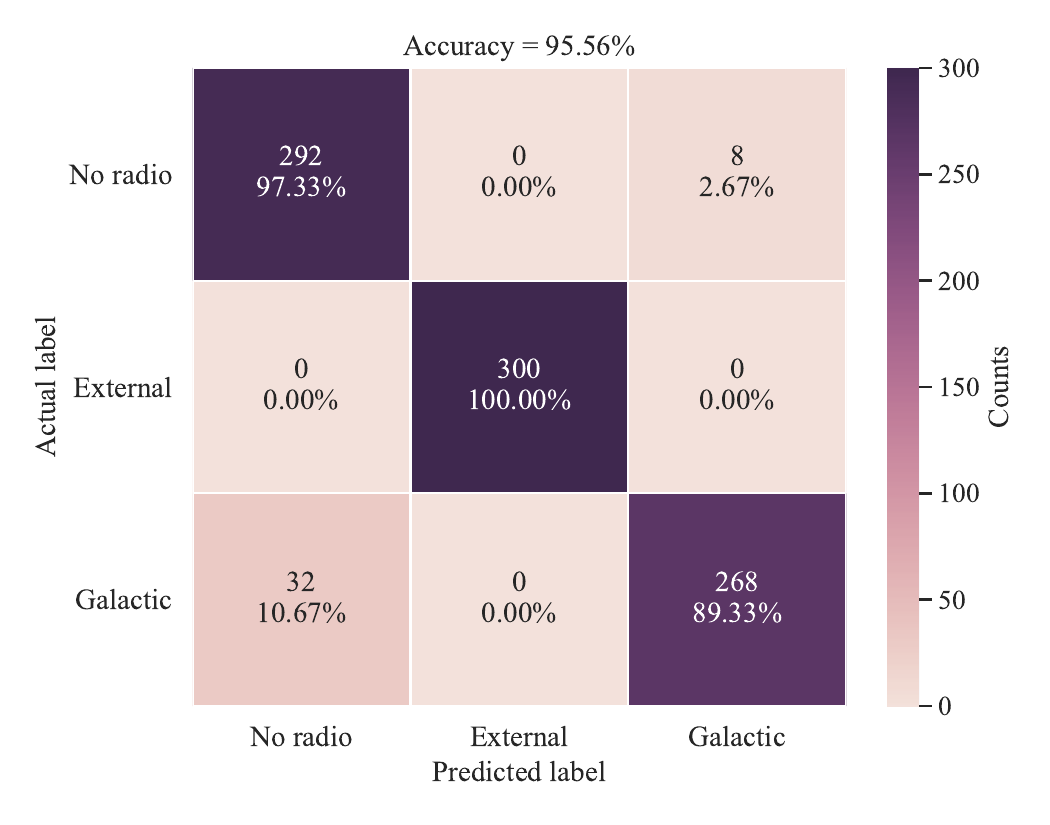}
    \end{minipage}\hfill
    \begin{minipage}{0.48\textwidth}
    \centering
    \textbf{Global signal}
       \includegraphics[scale=0.5]{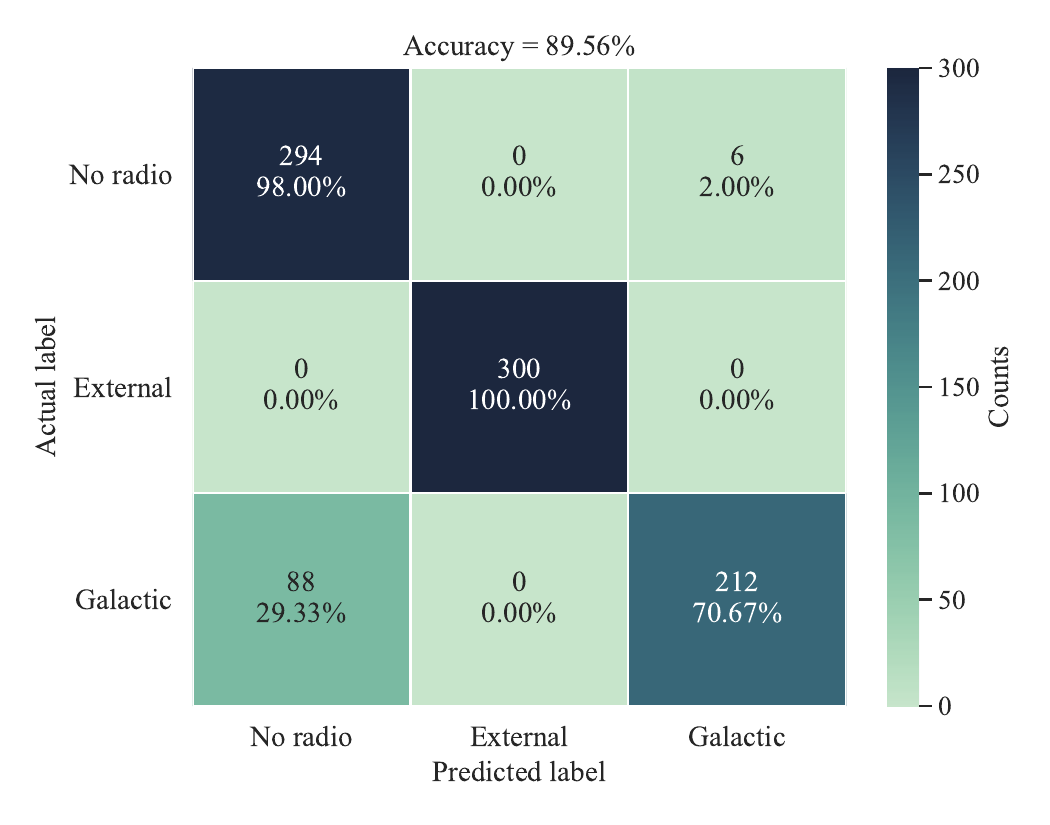}
    \end{minipage}
    \caption{\textbf{Left panel:} The confusion matrix shows the performance of classifying various radio backgrounds given the 21 cm power spectrum without any observational effects. The model classes were `No radio' (the standard astrophysical scenario without any excess radio background, i.e., CMB only), external radio models, and galactic radio models. The percentages add to 100 in each row, which corresponds to a particular true model class. Each column corresponds to a particular model class as predicted by the classifier. \textbf{Right panel:} A similar confusion matrix, but showing the performance of the classification procedure based on the global 21 cm signal (without any observational noise).}\label{fig:fig1}
\end{figure*}

As noted above, in this work we employ standard astrophysical models (CMB background only) along with both excess radio background models (galactic and external). We train an ANN classifier to distinguish the type of radio background present in the signal. The test data set consists of a total of 900 models, with 300 models from each category. To select the excess radio models for the test data set, we follow this approach:

\begin{enumerate}
    \item In the test data set, we excluded the most extreme EDGES-motivated models \citep[as were considered, e.g., in][]{sikder} and instead focused on models with moderate excess radio backgrounds, since such backgrounds might be realized in nature with a population of radio galaxies that is relatively similar to that expected in a normal astrophysical scenario. To this end, we set the maximum limit for $f_{\rm{Radio}}$ in the galactic radio background models, $f_{\rm{Radio, max}}$, to 300. 
    \item To make it more difficult and realistic for the classifier, we wanted to ensure that the test data sets with the two different radio backgrounds have a similar range of global signals. Thus, we set a maximum depth ($T_{\rm{21, max}}$) of the global signal. To find this value quickly, we identified galactic models for which $f_{\star} > 0.35$, $V_{\rm{c}}< 10$ km s$^{-1}$ and $f_{\rm{Radio}} < f_{\rm{Radio, max}}$. These criteria for $f_{\star}$ and $V_{\rm{c}}$ were used to select models with efficient early star formation, leading to efficient Ly$\alpha$ coupling, which results in deeper absorption troughs in the global signal. 
    \item We randomly selected 300 models from a subset (897 models) of the entire galactic models, for which $f_{\rm{Ratio}} < f_{\rm{Radio, max}}$ and the absorption trough in the global signal is shallower than $T_{\rm{21, max}}$.
    \item Next, we identified the max value of $A_{\rm{r}}$, $A_{\rm{r, max}}$, of the external models that gives a global signal absorption trough just under $T_{\rm{21, max}}$. For this, we considered only external models where $f_{\rm{X}} < 0.001$, giving very low X-ray heating and thus strong absorption. 
    \item Finally, we randomly selected 300 models from a subset (3070 models) of the entire data set of external radio models, for which $A_{\rm{r}} < A_{\rm{r, max}}$ and the absorption trough in the global signal is shallower than $T_{\rm{21, max}}$.
\end{enumerate}

Having thus selected the test data, we create a training data set comprising 8,893 models, including 2,893 models with the CMB-only background, 3,000 models with an excess external radio background, and 3,000 models with an excess galactic radio background. These were chosen randomly out of our full sets of the various models (after excluding the models in the test data). Thus, the training data had roughly the same number of models of each class, and covered their full ranges, so the ANN did not know at the training stage that the test data of the two excess radio models would be constrained to have similar global signal ranges. 

In Fig.~\ref{fig:test_models} we present the envelopes of the global signal and power
spectrum for the test data of the three different classes of models
(CMB only, external, and galactic radio backgrounds). The models in the test data set were excluded from the training data set, meaning that the ANN classifier did not encounter these test models during hyperparameter tuning and training. Therefore, the performance analysis using the test data set provides a reasonable estimate of the network's prediction accuracy. For classification, we label the models as $``0"$ for CMB only (no excess radio), $``1"$ for external excess radio models, and $``2"$ for galactic excess radio models.

To construct an ANN classifier that uses the 21 cm power spectrum in 30 $z$ and 32 $k$-bins as the input, we employ a five-layer MLP (input layer, three hidden layers and output layer), with 134 neurons in each hidden layer. The activation function for the hidden layers is chosen to be a \textit{rectified linear unit} function \citep[ReLU,][]{nair2010rectified}, and a stochastic gradient-based optimizer, namely the \textit{adam} optimizer \citep{adam_optimizer}, is applied for weight optimization. For the network that uses the global signal in 30 $z$-bins as input to predict the type of radio background, we use a single hidden layer with 143 neurons.  This network uses the \textit{logistic sigmoid} function \citep{Han1995TheIO} as the activation function and the same weight optimizer as the power spectrum-based network. Due to the stochastic nature of the training process, different runs of the same model on the same data set can produce slightly different results. To ensure reliability, we train each ANN classifier 40 times with a fixed network architecture, as determined after hyperparameter tuning, and calculate the mode of predictions across these runs to evaluate the final classification accuracy.

The performance of our classification procedure is shown using confusion matrices in Fig.~\ref{fig:fig1}. The left panel of Fig.~\ref{fig:fig1} presents the classification results for various radio backgrounds based on the 21 cm power spectrum without observational noise. Among 300 models with no excess radio background, only $2.7\%$ were incorrectly classified as having a galactic radio background. For 300 true external radio models, there were no misclassifications. However, for true galactic radio models, $11\%$ were incorrectly classified as having no excess radio. The overall classification accuracy using the 21 cm power spectrum across a wide range of redshifts and wavenumbers was $96\%$. The right panel of Fig.~\ref{fig:fig1} shows the classification performance based on the global signal. For true standard astrophysical models, only 2.0\% of models were misclassified as having a galactic radio background, while for true external models, the ANN classifier achieved $100\%$ accuracy. For true galactic radio models, the network misclassified $29\%$  as having no excess radio. The overall classification accuracy in this case was $90\%$. 

Even when we use a data set without observational noise, there is still some effective uncertainty due to the limited sampling of the large multi-dimensional parameter space of each model. The higher accuracy in using the power spectrum is likely due to the greater degree of information that is available in this case. It is evident that models with an external radio excess can be classified with significantly higher accuracy ($100\%$ based on either the power spectrum or global signal). However, a significant fraction of galactic models were misclassified as having no excess radio. This is likely because in the galactic radio model, the radio emission follows the overall redshift evolution of star formation, like other radiation outputs that are also present in the standard astrophysical model. On the other hand, the external radio excess model considered here (see equation~\ref{eq:Texternal}) may not be of astrophysical origin, does not depend on astrophysical parameters of galaxies like the other radio model, and has a substantially different redshift evolution (as it is especially enhanced at high redshifts); it can therefore be more easily identified. We note that the results are driven by the detailed shapes (with respect to redshift and, for the power spectrum, wavenumber), not just overall signal amplitudes. For example, in the case of the global signal, 60\% of test-data galactic models fall within the maximum depth of standard (no radio) models (164~mK depth), along with 21\% of external models. The misclassification rates are much lower than these fractions. In the case of the power spectrum, 93\% of test-data galactic models fall within the maximum height of standard models ($1.4\times 10^3$~mK$^2$), along with 40\% of external models. We also note that the mis-classification of Galactic as No radio occurs much more frequently than the reverse; this asymmetry is likely affected by the test data for the Galactic models covering only a part of the range of the Galactic training data, while the classifier effectively compares a given model to the model class as characterized by the entire training set. We further explore the dependence on amplitude in Section~\ref{sec:signal_amplitude} below.

These results indicate that both ANN classifiers could potentially distinguish external radio excess from galactic radio background models and standard astrophysical models if such an excess is present in the observed 21 cm power spectrum and global 21 cm signal. We note that, up to this point, we have used the simulated global 21 cm signal and power spectrum without any observational effects for training and testing the classifiers. In a more realistic observational scenario, several challenges arise. In the next subsection (\ref{sec:classification_noisy_data}), we present our analysis based on mock realistic data sets.

\subsection{Classification using realistic mock data sets}\label{sec:classification_noisy_data}

\begin{figure*}
    \centering
    \begin{minipage}{0.4\textwidth}
    \centering
    \hspace{1.5cm}
    \textbf{Power spectrum}
       \includegraphics[scale=0.5]{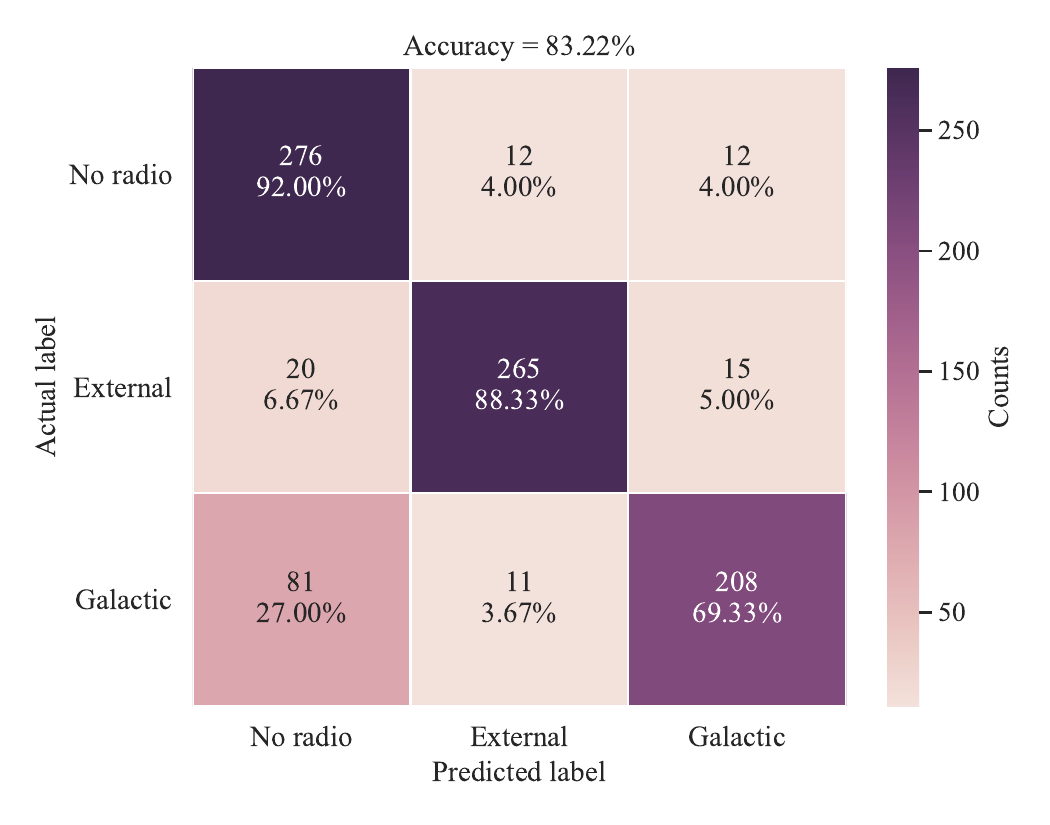}
    \end{minipage}\hfill
    \begin{minipage}{0.48\textwidth}
    \centering
    \textbf{Global signal}
       \includegraphics[scale=0.5]{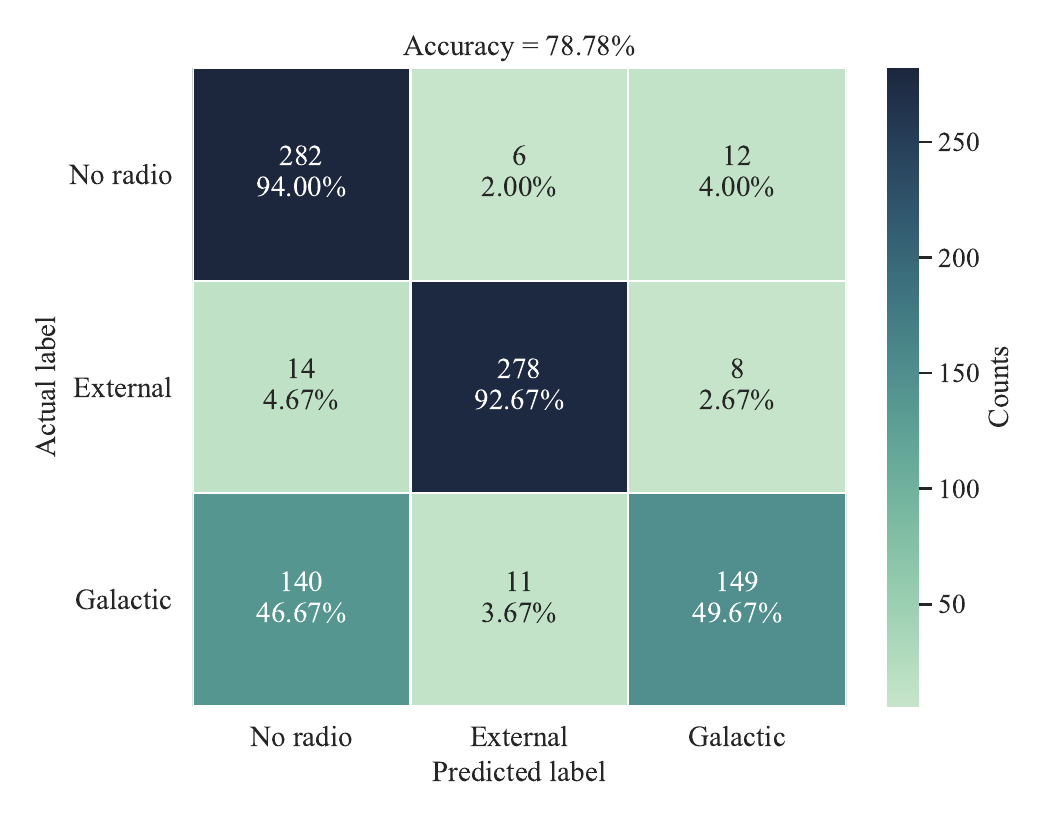}
    \end{minipage}
    \caption{Confusion matrices, using the same setup and notation as in Fig.~\ref{fig:fig1}, but here the data also include simulated observational noise. For the power spectrum we assume mock SKA 21 cm power spectra (which include thermal noise and other observational effects, see text). For the global signal, we include random Gaussian noise, for a relatively low noise level: $\mu = 0$ and $\sigma = 2.5$ mK; the results for additional noise levels ($\sigma = 17$~mK and 25~mK) are shown in Fig.~\ref{fig:confusion_matrix_gs_noise}.}\label{fig:fig1_SKA_noise1}
\end{figure*}

To assess the performance of the classification procedure in a more realistic scenario, we trained an ANN classifier using mock SKA 21 cm power spectra. We adopted the same activation function and weight optimizer as in the noiseless case, but here we standardized the mock SKA power spectra data set using the \textit{QuantileTransformer} scheme and utilized three hidden layers with 200 neurons each. The left panel of Fig.~\ref{fig:fig1_SKA_noise1} displays the confusion matrix summarizing the classification results based on the mock SKA 21 cm power spectra. Of the 300 models with no excess radio background, $8\%$ were misclassified as having an excess radio background. For the 300 models with an external radio background,  $88\%$ were correctly classified, while 69\% of true galactic radio models were accurately predicted to have the correct type of radio excess. We found that including various expected observational effects in the power spectra led to a lower overall classification accuracy of $83\%$, compared to $96\%$ in the case without any observational noise. The reduction in accuracy can be attributed to the observational effects as also reflected in having fewer $k$ and $z$ bins available for the 21 cm power spectra with mock SKA features. While the galactic radio case is most affected (in terms of the change in the percentages), the external radio case is also significantly affected, as the SKA observational effects are strongest at the highest redshifts.

For a realistic assessment of our classification procedure using the global signal, we added random Gaussian noise to the ideal global 21 cm signal from the simulation, in order to account for the sensitivity of various global signal telescopes (and analysis pipelines). We then constructed an ANN classifier using a similar architecture as the network that used the global signal without observational noise. The right panel of Fig.~\ref{fig:fig1_SKA_noise1} presents the classification performance for one of three different noise levels—specifically, an optimistic level of random Gaussian noise (with $\sigma=2.5$ mK and $\mu = 0$) added to the simulated sky-averaged signal. The classifier in this case achieved an overall accuracy of $79\%$ in identifying the type of radio background. Since the classifier determines the radio background type based on the shape and amplitude of the global signal, the added random Gaussian noise complicates the learning algorithm’s ability to track the shape and altered amplitude, resulting in reduced classification accuracy. A complementary analysis of higher Gaussian noise levels ($\sigma = $ 17 and 25 mK) is presented in Appendix A (see Fig.~\ref{fig:confusion_matrix_gs_noise}). These stronger noise cases yield substantially lower classification accuracies for the Galactic and External radio classes, though the standard model class remains relatively robust. This highlights the fact that inferring the type of radio background from data sets of existing telescopes will be challenging unless the observed signal shows a significantly enhanced radio background compared to the standard astrophysical model.

%Higher Gaussian noise levels with $\sigma$ values of 17 and 25 mK yielded significantly lower classification accuracies due to substantial distortion in the shape and amplitude of the global signal (see Fig.~\ref{fig:confusion_matrix_gs_noise}). The standard model class still achieves $91\%$ accuracy, but the Galactic radio class is around $40\%$ and even the External radio class is down to $\sim 60\%$. The classification is mainly enabled by the distinct behavior at high redshifts, as we discuss later; at the high-redshift end, the signal amplitude is relatively low, making it more susceptible to the added noise. Inferring the type of radio background from datasets of existing telescopes will be challenging unless the observed signal shows a significantly enhanced radio background compared to the standard astrophysical model. 

\subsection{The dependence of radio background classification on global signal amplitude}\label{sec:signal_amplitude}

\begin{figure}
\centering
\includegraphics[scale=0.5]{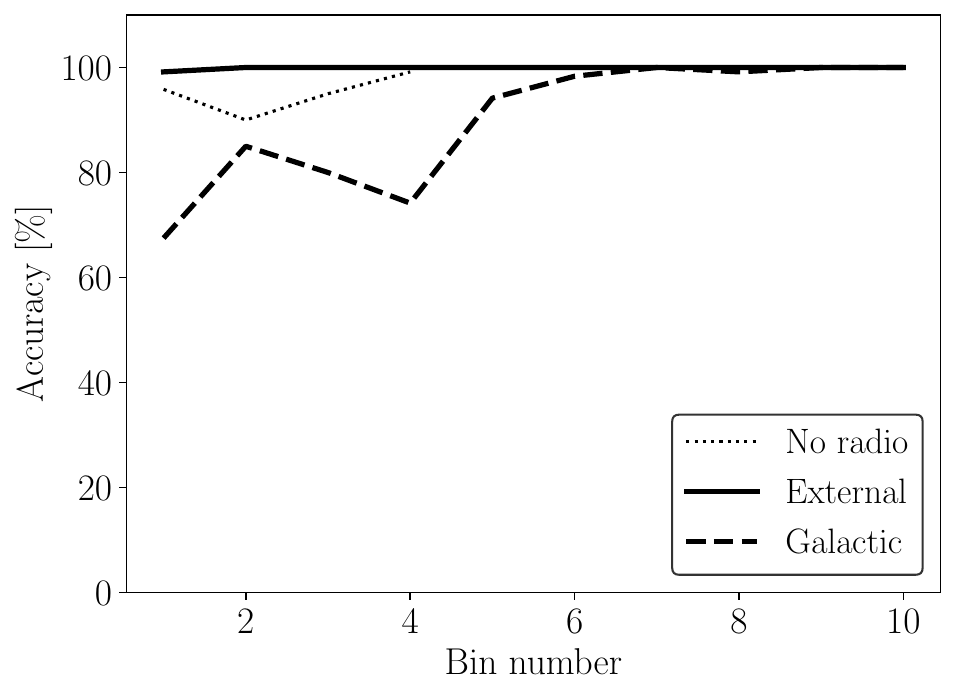}
\caption{Classification accuracy as a function of the amplitude of the global signal absorption trough (without observational noise). The higher the bin number, the higher is the signal amplitude (see Table~\ref{tab:amplitudes_bins} for the amplitude range of each bin).}
\label{fig:accuracy_vs_bin_number}
\end{figure}

\begin{table}
\centering
\begin{tabular}{lcc} 
\hline
Bin number ~    & Range in mK  & $\#$ of samples in test set \\ 
\hline\hline
$1$  & $[0, -50]$     & $120+120+120$ \\
$2$  & $[-50, -100]$    & $120+120+120$ \\
$3$  & $[-100, -150]$     & $120+120+120$ \\
$4$  & $[-150, -200]$      & $120+120+120$ \\
$5$  & $[-200, -250]$    & $120+120$\\
$6$  & $[-250, -300]$     & $120+120$\\
$7$  & $[-300, -350]$   & $120+120$\\
$8$  & $[-350, -400]$    & $120+120$\\
$9$  & $[-400,-450]$    & $120+120$\\
$10$ & $[-450,-500]$    & $120+120$\\

\hline
\end{tabular}
\caption{Amplitude ranges (in mK) of the absorption troughs considered in 10 bins. For each amplitude bin, we selected 120 samples from each class, resulting in a total of 360 samples per test data set. Since the maximum absorption trough depth for the standard (No radio) model is approximately $-165$ mK, for bins greater than 4 we only use external and galactic models with maximum absorption depth corresponding to the particular bin.}\label{tab:amplitudes_bins}

\end{table}

To explore how the depth of the global signal absorption trough affects the classification accuracy, we divided the data (for the case without observational noise) into 10 separate bins based on the trough depth. The amplitude ranges for each bin are listed in Table~\ref{tab:amplitudes_bins}. For this analysis, we set an upper limit of $-500$ mK for the absorption amplitude, aligning with the best-fit amplitude suggested by the reported EDGES detection \citep{bowman18}. Each bin represents a distinct test data set containing 360 samples, composed of 120 randomly selected global signal samples from each class, where the absorption troughs fall within the specific range of the bin (except that bins that are beyond the range of No radio models do not contain this class). These test data sets are then applied to the trained classifier to evaluate the classification accuracy. Fig.~\ref{fig:accuracy_vs_bin_number} shows the variation in classification accuracy as a function of the global signal amplitude (where a higher bin number indicates a deeper absorption trough). Deeper absorption troughs lead to more accurate classification of the excess radio models. For the galactic model in particular, the accuracy for bins five and up is significantly higher than for the first four bins, which suggests that once the absorption depth exceeds values expected from standard astrophysical scenarios, the depth itself becomes a key factor in distinguishing between excess radio models and standard models. However, even for the first 4 bins, the classification accuracy is fairly high for all model classes, especially for the external radio background. We note that Fig.~\ref{fig:accuracy_vs_bin_number} does not exactly correspond to the results in section~4.1 since here the various models were selected from the full data sets with only a restriction on the amplitude.

\begin{figure*}
    \centering
    \includegraphics[scale=0.5]{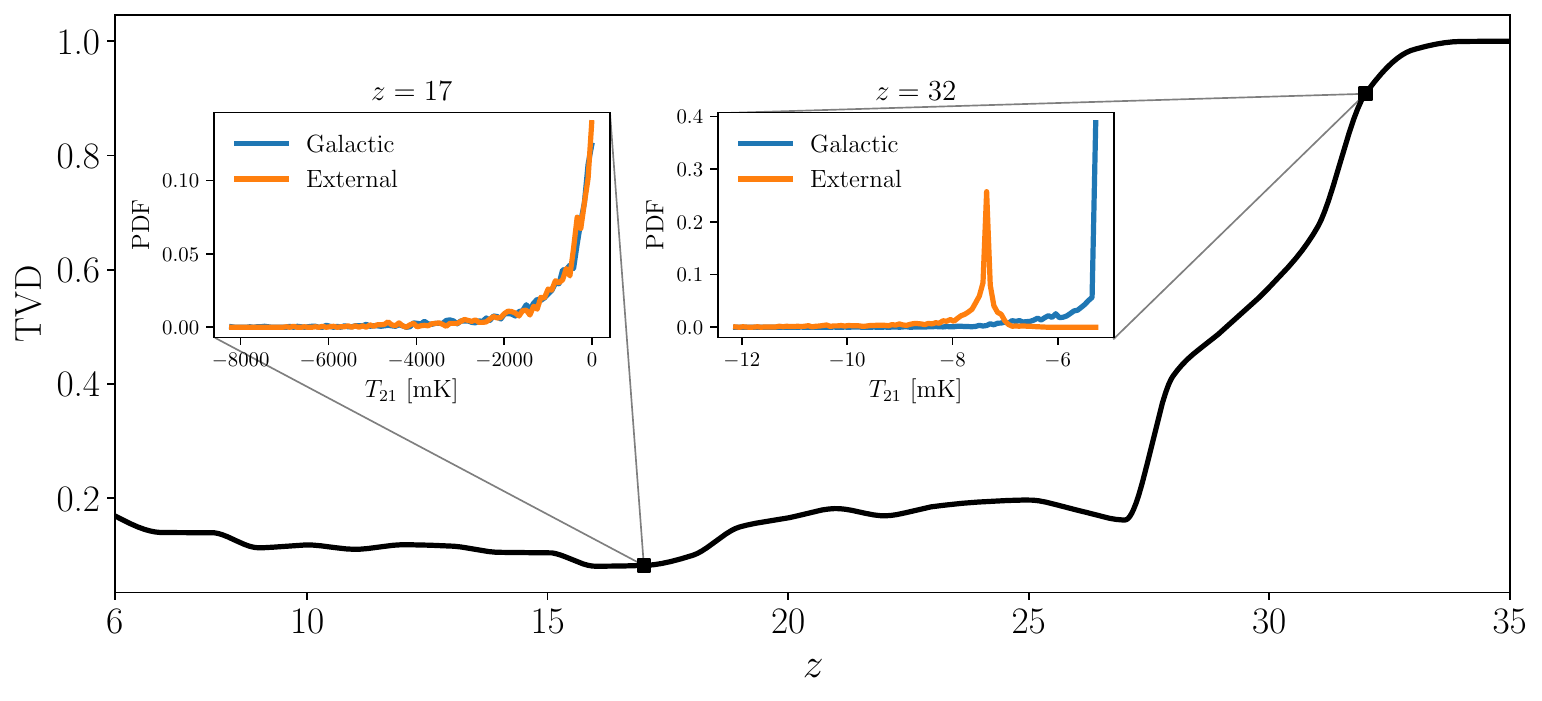}
    \caption{The total variation distance (TVD) as a function of $z$ between the normalized PDFs of the global signals (without observational noise) of galactic and external models in the training data set. TVD quantifies the difference between two probability distributions (see text). The inset plots show the normalized PDFs of the sky averaged 21 cm brightness temperatures from galactic and external models (all models in the training sets) at $z = 17$ and $32$. At lower redshifts ($<27$), the value of the TVD is low compared to that at high redshifts, indicating that the difference between the normalized PDFs from galactic and external samples is much larger at high redshifts. Thus, the high-redshift behavior of external models is likely the reason behind the accurate classification of those models relative to the other two models, irrespective of the overall global signal amplitude.}
    \label{fig:l1_norm_z}
\end{figure*}

To help understand why the classification of the external radio background is so effective, we conducted an additional analysis. We selected both external and galactic radio models from our training data set and statistically evaluated the differences in their brightness temperature distributions at each redshift. We quantified this distinction using a statistical distance metric known as the total variation distance (TVD), which measures the maximum difference between two distributions. The TVD is related to the L1-norm and defined by the equation~\citep{LevinPeresWilmer2006}:
 \begin{equation}
     \mathrm{TVD} = \frac{1}{2}\sum_i | P_i - Q_i| \ ,
 \end{equation}
where $P$ and $Q$ represent the two distributions, and $i$ is the bin number. A TVD value of zero indicates that the two distributions are identical, while a TVD value of 1 suggests complete dissimilarity between the normalized distributions. Higher TVD values reflect greater differences between the distributions. We calculated the TVD at each redshift, and its evolution with redshift is shown in Fig.~\ref{fig:l1_norm_z}. Below redshift 27, the TVD hovers around or below 0.2. However, beyond this point, the TVD rapidly increases with redshift, indicating significant differences between the temperature distributions of the external and galactic models at redshifts greater than 27. The inset plots in Fig.~\ref{fig:l1_norm_z} compare the normalized temperature distributions of the galactic and external models at at $z = 17$ and $32$. While the distributions are similar at $z = 17$, they diverge significantly at $z = 32$. These high redshift differences could potentially account for the accurate classification of the external radio background, regardless of the global signal amplitude. In the next subsection we further explore various scenarios by selecting distinct redshift ranges for training and subsequently calculating accuracy scores for the same test data set.

\subsection{Classifications using REACH and SARAS~3 bands}

\begin{figure}
\centering
\includegraphics[scale=0.5]{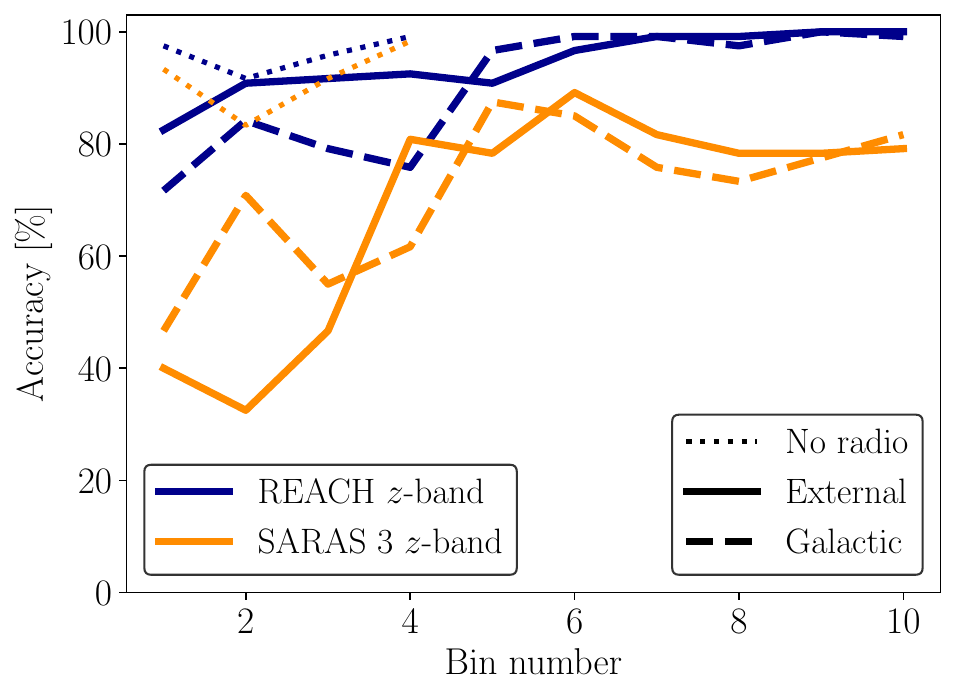}
\caption{Classification accuracy as a function of the amplitude of the global signal absorption trough. Here we consider only the accessible redshift range of the REACH or SARAS~3 experiment when we train the classifier for excess radio background model classification. The bins refer to Table~\ref{tab:amplitudes_bins}.}
\label{fig:accuracy_vs_bin_number_saras3_reach}
\end{figure}

In this subsection, we consider the redshift intervals covered by global signal experiments like REACH ($z: 7-28$) \citep{de_lora2022} and SARAS~3 ($z: 15-25$) \citep{bevins2022}. We train the classifier using simulated global signal data specific to these redshift ranges and then replicate the analysis performed in  \ref{sec:signal_amplitude}. For this analysis, we utilize the simulated global signal data without any added Gaussian random noise. Fig.~\ref{fig:accuracy_vs_bin_number_saras3_reach} shows the variation in accuracy with the depth of the absorption troughs for both the REACH (dark blue lines) and SARAS~3 (dark orange lines) redshift bands.

For the REACH band, the classification accuracy is generally similar to that shown found for the full redshift range in Fig.~\ref{fig:accuracy_vs_bin_number}. The only significant difference occurs for the external radio model. This reaffirms the notion highlighted in the previous section that very high-redshift signal information is important for accurately distinguishing external models from the other two categories. Still, the accuracy for classifying the various model classes is high, even for low absorption amplitudes. This indicates that global signal observations from REACH-like experiments have the potential to detect the presence or absence of an external radio background in the 21 cm signal. The narrower redshift range of the SARAS~3 band results in lower classification accuracies. Again, the largest effect is on the External radio model, so that its accuracy becomes similar to that of the Galactic radio model. At low amplitudes that are within the range of standard astrophysical models, the excess radio models have a classification accuracy comparable to 50\% with the SARAS~3 redshift range, while the accuracy is still high ($\sim 80\%$) at higher absorption amplitudes.

\subsection{$k$-fold cross-validation}\label{sec: cross-validation}

\begin{table*}
\centering
\begin{tabular}{lccccccc} 
\hline
 signal & Fold 1 & Fold 2 & Fold 3 & Fold 4 & Fold 5 & Fold 6 & Mean \\ 
\hline\hline
Power spectrum & 98.40 & 98.0 & 98.0 & 98.13 & 97.73 & 98.27 & 98.09 \\
Global signal  & 97.20 & 96.80 & 97.4 & 97.0 & 96.40 & 96.40 & 96.87 \\

\hline
\end{tabular}
\caption{Accuracy score ($\%$) in predicting the type of the radio background based on the power spectrum and the global signal  for each case of 6-fold cross-validation.}
\label{tab:table_CV}
\end{table*}

\begin{figure*}
    \centering
    \begin{minipage}{0.4\textwidth}
    \centering
    \hspace{1.5cm}
    \textbf{Power spectrum}
       \includegraphics[scale=0.5]{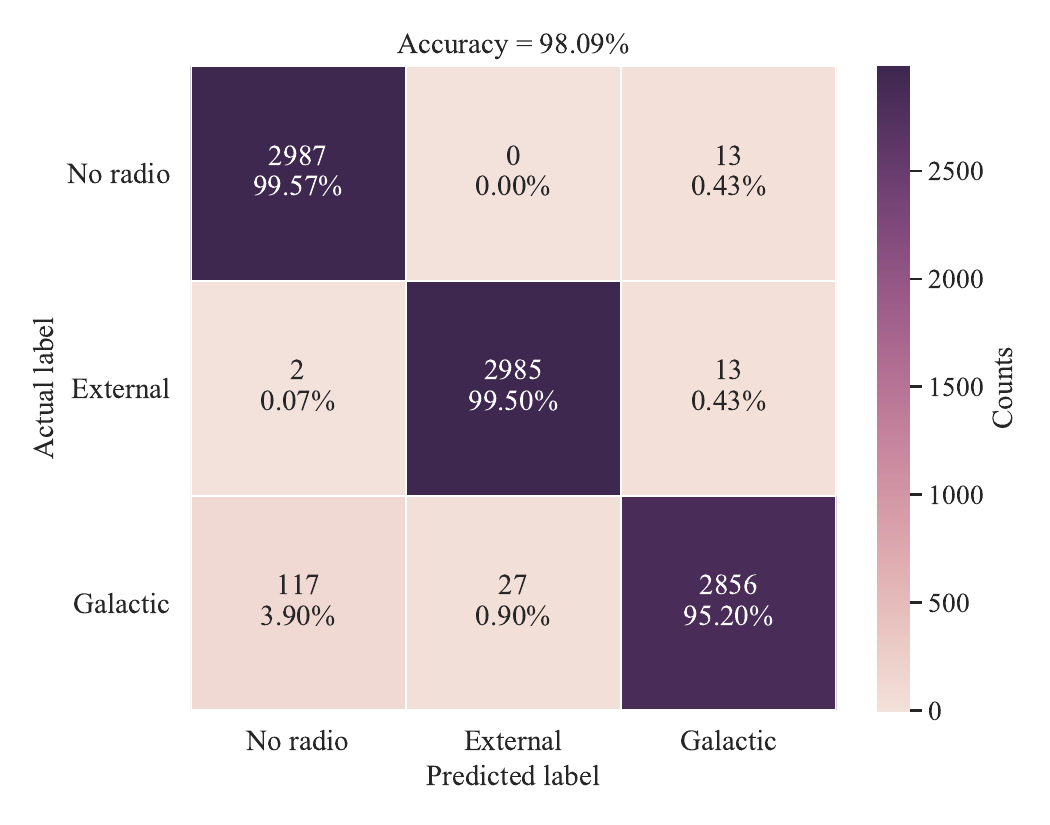}
    \end{minipage}\hfill
    \begin{minipage}{0.48\textwidth}
    \centering
    \textbf{Global signal}
       \includegraphics[scale=0.5]{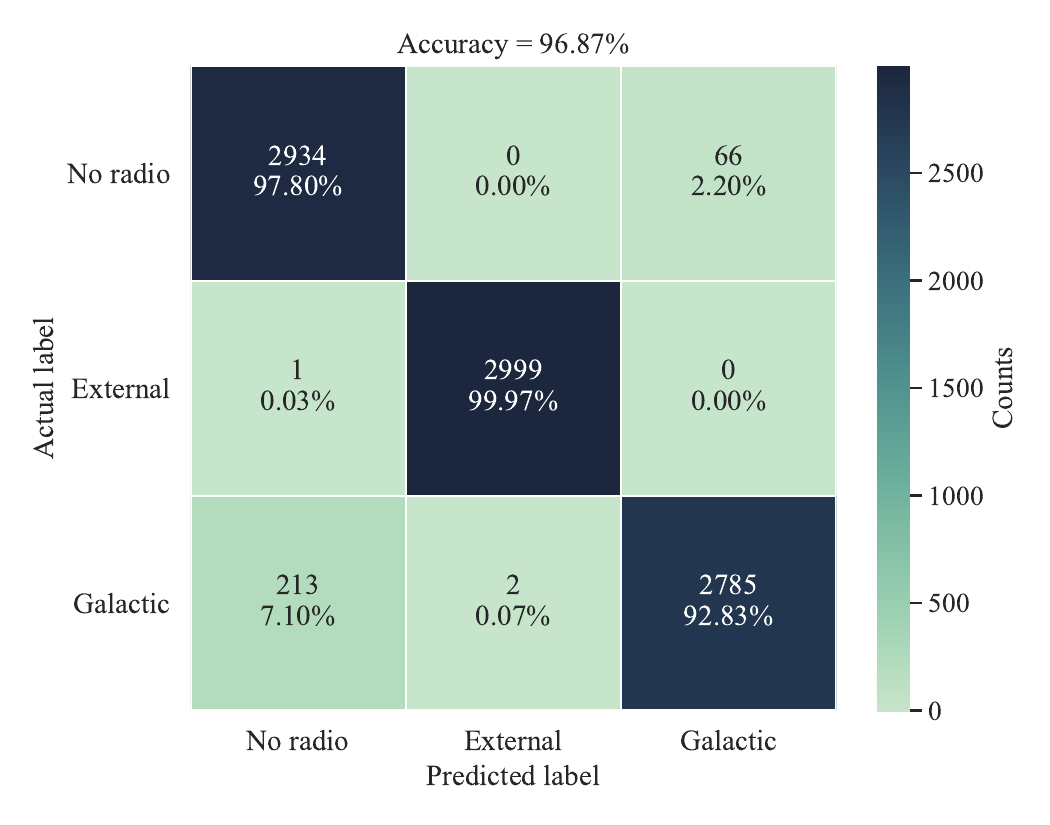}
    \end{minipage}
    \caption{Confusion matrices, using the same setup and notation as in Fig.~\ref{fig:fig1}, but here we show the combined confusion matrices from all six folds.}\label{fig:fig1_CV}
\end{figure*}

So far, we have presented our analysis based on a specific test data set, constructed using the criteria outlined in Section \ref{sec:classification_clean}. To further evaluate the performance of our classification model, we adopt a more statistically rigorous approach by implementing $k$-fold cross-validation. In this method, the training data set is divided into $k$ subsets (or ``folds"). In each iteration, one subset is used as the validation (test) set while the model is trained on the remaining $k-1$ subsets. This process is repeated $k$ times, with each subset serving as the test set exactly once. By comparing the outcomes across all iterations, we can assess the consistency and reliability of the classifier. For the cross-validation procedure, we construct a comprehensive training data set by randomly selecting 3000 samples from each of the three categories, resulting in a total of 9000 samples. We choose $k=6$ for our analysis. In each fold, the validation set consists of 1500 samples (500 from each category), representing approximately $17\%$ of the entire data set. Fig. \ref{fig:fig1_CV} displays the combined confusion matrices from all six folds: the left panel shows the classification performance based on the 21 cm power spectrum, while the right panel shows the results based on the global signal. Among the 3000 models with no radio excess, only 13 were incorrectly classified as having an excess galactic radio background when using the power spectrum. For the 3000 external radio models, the number of misclassifications is 15. In the case of true galactic radio background models, $3.90\%$ were misclassified as having no radio excess, and $0.90\%$ as having an external radio background. Using the global signal as input, only $2.20\%$ of 3000 no-radio-excess models were misclassified as having a galactic radio background. Among the true external radio models, just 1 model was misclassified. For the galactic radio background category, $7.10\%$ of the models were misclassified as having no radio excess, and $0.07\%$ as having an external radio background. Table \ref{tab:table_CV} reports the overall accuracy for each fold, separately for classifications based on the power spectrum and the global signal. The average accuracy across all folds is $98.09\%$ for the power spectrum and $96.87\%$ for the global signal. These results demonstrate the robustness and consistency of our classification model when applied to the entire training data set. Unlike the earlier analysis shown in Fig. \ref{fig:fig1}, which used a fixed test set of 900 models, cross-validation leverages the entire data set of 9000 models by iteratively training and testing on different partitions. This exposes the model to a broader variety of examples, enhancing its generalization capability and improving its accuracy. Overall, cross-validation provides a more reliable estimate of model performance, confirming the trends observed in Fig. \ref{fig:fig1}, but with greater precision and confidence.

\section{Mapping 21 cm observables: the global signal and the 21 cm power spectrum}\label{sec:mapping_21cm_observables}

\begin{figure}
    \centering
    \includegraphics[width=0.96\linewidth]{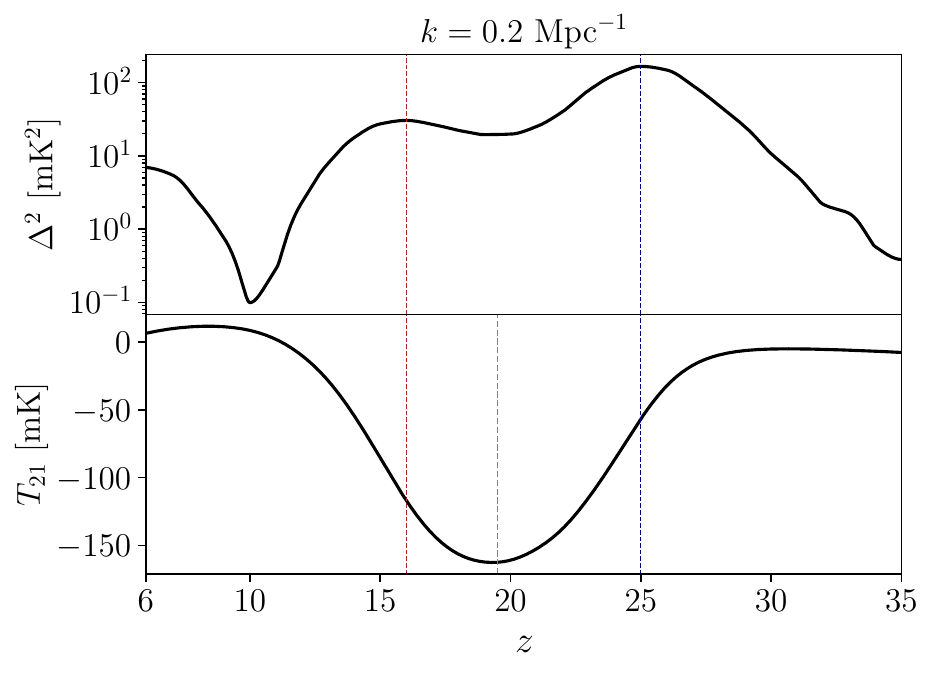}
    \caption{An example power spectrum at $k=0.2$ Mpc$^{-1}$ (top panel) and corresponding global signal (bottom panel) of standard astrophysical model from the full data set. The dashed blue and red line represent the redshifts at which Ly$\alpha$ and heating peak appear in the power spectrum, respectively. The dashed gray line represent the global signal absorption trough.}
    \label{fig:example_T21_delta_21}
\end{figure}

While previous studies in the literature have utilized ANN to construct emulators for the power spectrum and global signals, aiming to estimate the astrophysical parameters of cosmic dawn and the EoR, here we first apply an ANN for the direct mapping of the two key observables in 21 cm cosmology: the global signal, which could potentially be observed using a single radio antenna, and the 21 cm power spectrum, a statistical measurement through radio interferometry. Traditionally, the emulation approach involves obtaining data for one of these observables and then employing these data to constrain the underlying astrophysical model. Subsequently, with the help of the determined astrophysical model, predictions can be made for the other observable, utilizing machine learning algorithms proposed in the existing literature. The novelty here lies in the capability of the proposed network to directly predict the feasible range of one observable when provided with observations of the other. This streamlined approach represents a potentially useful and efficient way to analyze 21 cm cosmology data from upcoming experiments.

Fig. \ref{fig:example_T21_delta_21} illustrates an example of the 21 cm power spectrum at $k=0.2$ Mpc$^{-1}$ (top panel) alongside the corresponding global signal (bottom panel), both derived from a standard astrophysical model within our full data set. At high redshifts ($z \sim 35$–$30$), corresponding to the pre-cosmic dawn era, density perturbations are already present. However, they do not yet generate a significant 21 cm signal due to weak Ly$\alpha$ coupling. During this epoch, the gas temperature ($T_{\rm{K}}$) remains close to the CMB temperature ($T_{\rm{CMB}}$), resulting in minimal contrast between the spin temperature ($T_{\rm{s}}$) and $T_{\rm{CMB}}$. Consequently, the global 21 cm signal remains nearly zero. As the Universe enters the cosmic dawn era ($z \sim 30$–$20$), the first generations of stars emit copious Ly$\alpha$ photons. These photons drive the Wouthuysen-Field effect, coupling $T_{\rm{s}}$ to $T_{\rm{K}}$. Due to adiabatic cooling, the gas is colder than the CMB ($T_{\rm{K}} < T_{\rm{CMB}}$), causing the global 21 cm signal to transition into absorption and become increasingly negative. Simultaneously, spatial variations in Ly$\alpha$ intensity—tracing the biased distribution of the first star-forming regions—introduce fluctuations in the 21 cm signal. This leads to a pronounced feature in the power spectrum, known as the Ly$\alpha$ peak,  appearing at $z = 25$ in this example (see Fig. \ref{fig:example_T21_delta_21}). Stronger global absorption tends to correlate with higher $\Delta^2$ values during this phase, as both phenomena trace the onset and progression of star formation. Following this period, during the heating era ($z \sim 20$–$10$), X-ray photons from early galaxies heat the intergalactic medium (IGM), gradually increasing $T_{\rm{K}}$. As the gas temperature rises towards and eventually exceeds $T_{\rm{CMB}}$, the global 21 cm absorption trough begins to diminish, and the signal moves back toward zero. In this phase, spatial fluctuations in X-ray heating dominate the power spectrum, producing the heating peak, which occurs at $z = 16$ in the example shown. The decline in the global signal amplitude during this period reflects the average heating of the IGM, while the heating peak in the power spectrum captures spatial variations in the temperature field. As heating progresses and becomes more uniform, temperature fluctuations saturate, causing $\Delta^2$ to decline. Subsequently, during the epoch of reionization, the growth of ionized regions introduces large-scale fluctuations in the neutral hydrogen fraction ($x_{\rm{HI}}$). These fluctuations give rise to a new peak in $\Delta^2$, typically occurring near mid-reionization. As ionizing UV photons from galaxies progressively ionize the IGM, reducing $x_{\rm{HI}}$, the global 21 cm signal continues to decline toward zero. Once reionization completes, both the large-scale 21 cm fluctuations and the global signal vanish. 

The example presented in Fig. \ref{fig:example_T21_delta_21} provides a qualitative illustration of the strong correlation between the global 21 cm signal ($T_{21}$) and the 21 cm power spectrum amplitude ($\Delta^2$) for a single astrophysical scenario. To go beyond this example and provide a more statistically robust quantification of the relationship between $T_{21}(z)$ and $\Delta^2(k, z)$, we next compute both the Pearson and Spearman correlation coefficients over all samples from the standard astrophysical model. These coefficients are evaluated across different redshifts and wavenumbers, offering insights into the correlation between the global 21 cm signal and its spatial fluctuations throughout cosmic history.

Pearson correlation measures the strength and direction of the linear relationship between two data sets. It's value ranges from $-1$ to $+1$ where 0 indicates no correlation. A coefficient of $+1$ or $-1$ implies an exact linear relationship, either positive or negative, respectively. In this context, a high Pearson correlation at a given ($k, z$) suggests that $T_{21}(z)$ and $\Delta^2(k, z)$ vary proportionally across samples. On the other hand, the Spearman correlation coefficient assesses the strength and direction of a monotonic relationship, which does not have to be linear. Like the Pearson coefficient, it ranges from $-1$ to $+1$, with $0$ indicating no correlation. Values of $+1$ or $-1$ indicate a perfect monotonic relationship. A high Spearman correlation means that as $T_{21}(z)$ increases (or decreases), $\Delta^2(k, z)$ consistently increases (or decreases) as well, even if the rate of change is not constant. Comparing both Pearson and Spearman correlations is important for distinguishing whether the relationship between these two 21 cm observables is strictly linear or generally monotonic which is crucial for understanding the physical connection between these 21 cm observables. 

\begin{figure}
    \centering
    \includegraphics[width=0.95\linewidth]{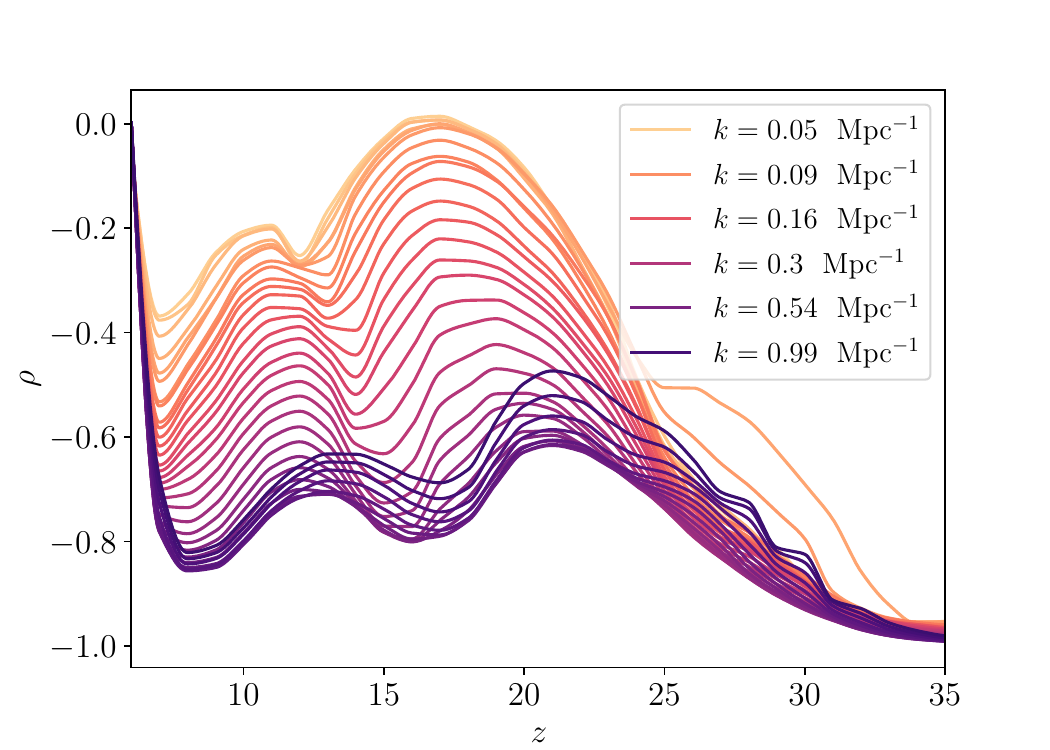}
    \includegraphics[width=0.95\linewidth]{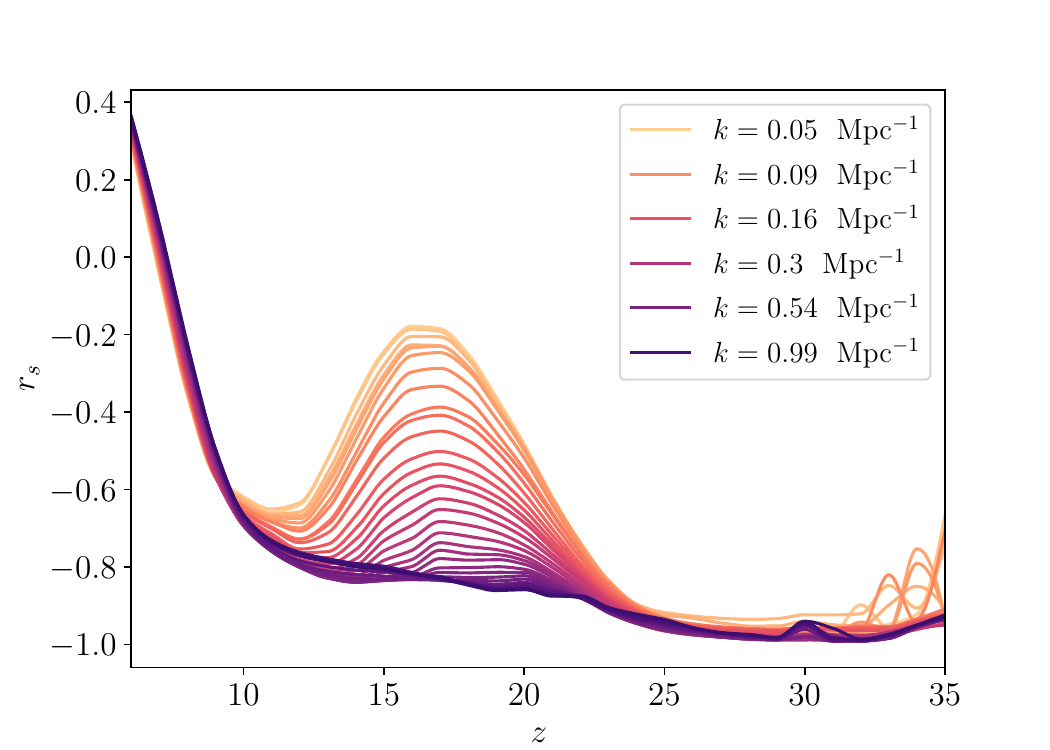}
    \caption{\textbf{Top panel:} Pearson coefficient ($\rho$) calculated at each $z$ and $k$ over all the samples of standard astrophysical model. $\rho$ is negative across all redshifts and $k$-modes.  This indicates a consistent inverse linear relationship between the global signal and power spectrum.  At higher redshifts, the correlation is the strongest (close to -1) for all $k$-modes. At intermediate redshifts,  the anti-correlation weakens.  Strong anti-correlation at all scales during high redshifts reflects the tight connection between $T_{21}$ and $\Delta^2$. \textbf{Bottom panel:} Spearman coefficient ($r_s$) calculated at each $z$ and $k$ over all the samples of standard astrophysical model.  At high redshifts, the spatial fluctuations in power spectrum at all scales strongly dictates the global signal. The persistence of Spearman anti-correlation at large scales and early redshifts suggests that $T_{21}$ and $\Delta^2$ maintain a strong monotonic relationship, even if not purely linear.}
    \label{fig:pearson_spearman}
\end{figure}

The top panel of Fig. \ref{fig:pearson_spearman} presents the Pearson correlation coefficient ($\rho$), calculated at each redshift $z$ and wavenumber $k$, using all samples from the standard astrophysical model. Across all redshifts and $k$-modes, $\rho$ is consistently negative, indicating an inverse relationship between the global signal, $T_{21}(z)$, and the power spectrum, $\Delta^2(k, z)$. However, this relationship is not necessarily perfectly linear at all times. At the highest redshifts, the anti-correlation is the strongest, with $\rho \sim -1$ across all $k$-modes. At intermediate redshifts, the anti-correlation weakens, and a scale-dependent trend emerges: small-scale modes (larger $k$) maintain stronger correlations with $T_{21}$ compared to large-scale modes (smaller $k$).

The bottom panel of Fig. \ref{fig:pearson_spearman} shows the Spearman correlation coefficient ($r_s$), similarly computed at each $z$ and $k$. It reveals a strong negative monotonic relationship between $T_{21}(z)$ and $\Delta^2(k, z)$ at high redshifts. As redshift decreases, this monotonic anti-correlation persists more prominently at small scales (large $k$), even when linearity (as shown by the Pearson coefficient) diminishes. The transition from strong anti-correlation at high redshift across all scales to weaker, scale-dependent correlations at lower redshifts suggests an evolution closely tied to the underlying astrophysical processes. The Spearman coefficient highlights the persistence of monotonic trends, reinforcing the presence of a systematic relationship even when strict linearity fades.

These correlations illustrate how closely the global signal and the power spectrum are connected—both linearly (Pearson) and monotonically (Spearman)—across redshift and spatial scales. This systematic behavior implies that an ANN, in principle, can exploit these correlations. For example, it can predict $\Delta^2(k, z)$ from $T_{21}(z)$ by leveraging the strong anti-correlation observed at high redshifts across all $k$-modes. Conversely, it may predict $T_{21}(z)$ from $\Delta^2(k, z)$ by integrating the persistent small-scale correlations at lower redshifts.

\subsection{Prediction of the global 21 cm signal}\label{sec:gs_prediction}

\subsubsection{Method}

\label{sec:A1}

Signal validation will become a crucial aspect of 21 cm data analysis once detections from multiple experiments become available. In particular, the consistency of the detections between interferometers and radiometers would be a strong sign of the true cosmological signal. Here, we develop a methodology to enable such validation. 

\begin{figure}
    \centering
    %\begin{minipage}{0.4\textwidth}
       \includegraphics[scale=0.45]{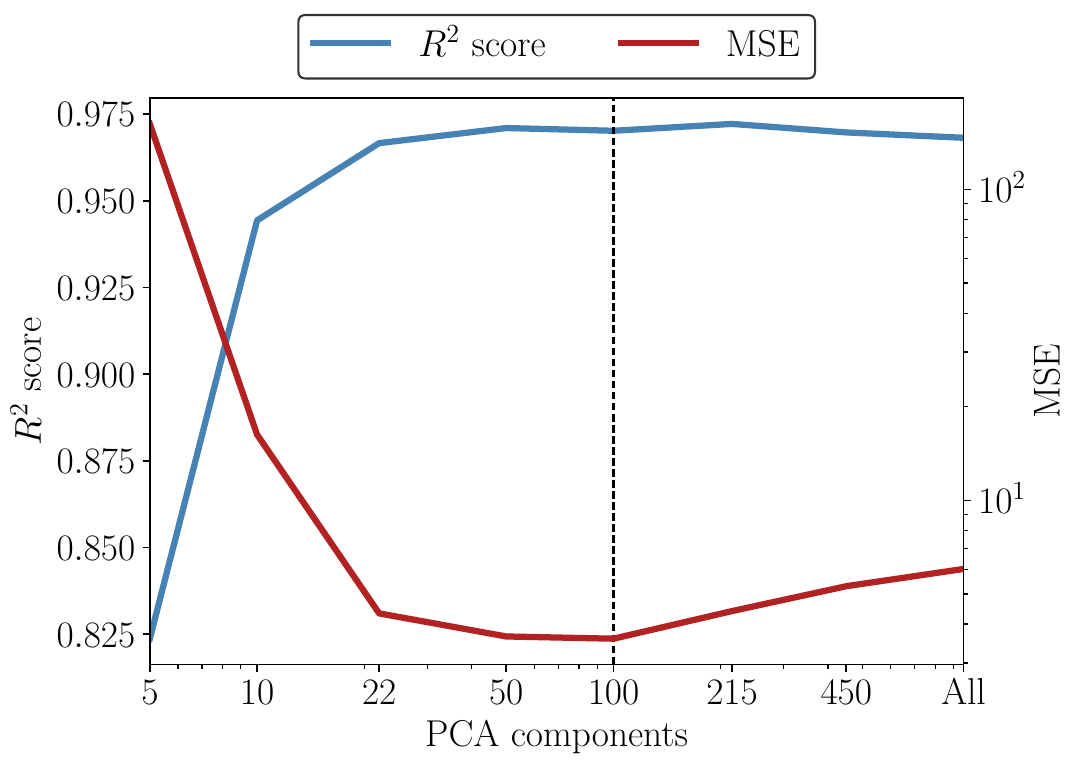}

    \caption{The impact of dimensionality reduction of the power spectrum data set in the context of predicting the global signal given the power spectrum spectrum without any observational effects. We show how the two regression performance estimators, $R^2$ score and MSE, vary with the number of PCA components in the learning model. The dashed vertical line shows the optimum number of components.}
    \label{fig:pca_component_gs_prediction}
\end{figure}

We construct an ANN to predict the global signal over a wide range of redshifts given the 21 cm power spectrum over the same redshift range. Here, as a proof of concept, we use only the data sets with standard astrophysical models (no excess radio, i.e., CMB only). The network consists of three hidden layers, each with 150 neurons, and employs similar data preprocessing steps as described previously, including standardization and PCA dimensionality reduction. To determine the optimal number of PCA components, we evaluate regression performance using the $R^2$ score and mean squared error (MSE) on the test data set. The $R^2$ score quantifies the goodness of fit so that an $R^2$ score $= 1$ implies a perfect fitting. The lower the MSE is, the better is the fitting quality. As shown in Fig. \ref{fig:pca_component_gs_prediction}, the optimal choice is 100 PCA components, which yields the highest $R^2$ score and lowest MSE. Accordingly, the network has a 100-dimensional input layer and a 30-dimensional output layer, predicting the global signal at 30 redshifts from $z = 6$ to 35. We use the ReLU as the activation function for the hidden layers and choose the \textit{adam} optimizer for weight optimization of the network. Once trained, the network is applied to the test data set for predictions.

% To do one of the preprocessing steps, i.e., standardization of the features, we use the \textit{StandardScaler} class from the \texttt{Scikit-learn} library \citep{scikit_learn}. After the standardization of the dataset, all the features have a zero mean and unit standard deviation. 
%The next data preprocessing step involves projecting the dataset into a lower dimensional space using PCA.

%To do one of the preprocessing steps, i.e., standardization of the features, we use the \textit{StandardScaler} class from the \texttt{Scikit-learn} library \citep{scikit_learn}. After the standardization of the dataset, all the features have a zero mean and unit standard deviation. 

In order to understand the performance of the prediction when using the 21 cm power spectrum with observational noise, we also train a network that can predict the global signal given the mock SKA 21 cm power spectrum. We use the same network architecture and the same preprocessing steps except that here we do not apply PCA to project the data set into a lower dimensional space, because in this case, we already use the 21 cm power spectrum binned over 8 redshifts and 5 $k$ values. Thus, the dimension of the input data set is already reduced by a factor of 20 compared to our noiseless data set.

\subsubsection{Performance analysis}

We evaluate the overall performance of the network in predicting the global 21 cm signal given the power spectrum using the following equation, where we define the error as the r.m.s.\ value of the difference between the predicted global signal ($T_{\mathrm{Predicted}}$) and the true global signal ($T_{\mathrm{True}}$) from the simulation with the same parameter set, normalized by the maximum amplitude of the true global signal:

\begin{equation}\label{eq:gs_prediction_error}
\mathrm{Error} = \frac{\sqrt{\mathrm{mean} \left[ (T_{\mathrm{True}}(z) - T_{\mathrm{Predicted}}(z))^2 \right]}}{\mathrm{max}|T_{\mathrm{True}}(z)|} \ . 
\end{equation}

\begin{figure}
    \centering
    \includegraphics[scale=0.5]{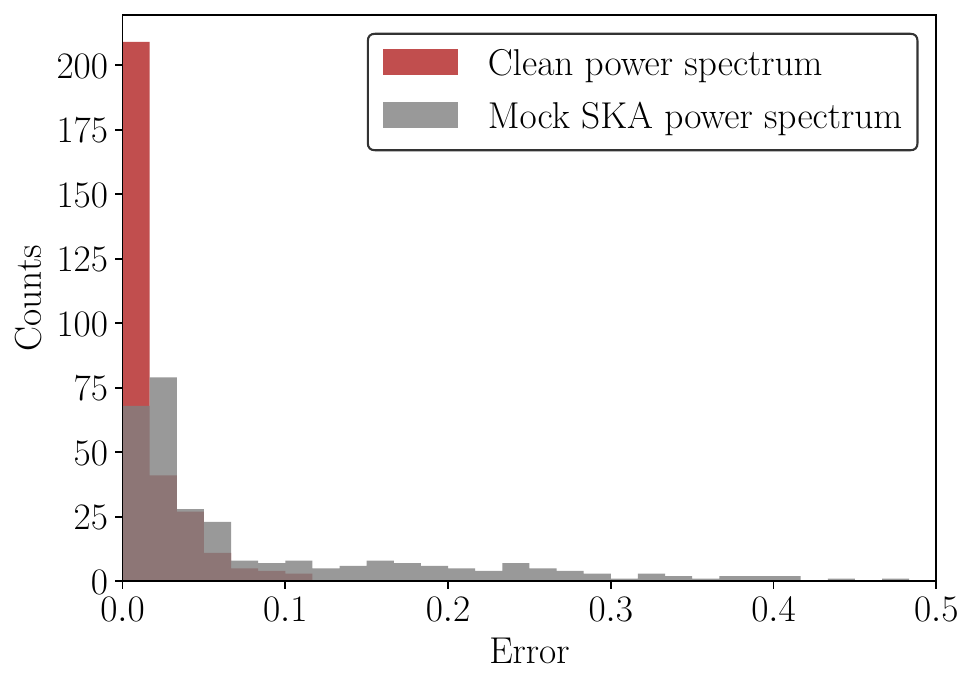}
    \caption{Histogram of the relative error in the global signal predicted based on a given power spectrum, where the relative error is defined by equation~\ref{eq:gs_prediction_error}. When the network is trained using the clean power spectrum, the median error in predicting the global signal is 0.0094, while for the network trained using a mock SKA power spectrum, the median error is 0.035. Here we use models that assume the standard astrophysical scenario (CMB only).}
    \label{fig:histogram_error_gs_prediction_clean_SKA}
\end{figure}

\begin{figure*}
    \centering
    \begin{minipage}{0.48\textwidth}
    \centering
    \textbf{Using the clean power spectra}
         \includegraphics[scale=0.35]{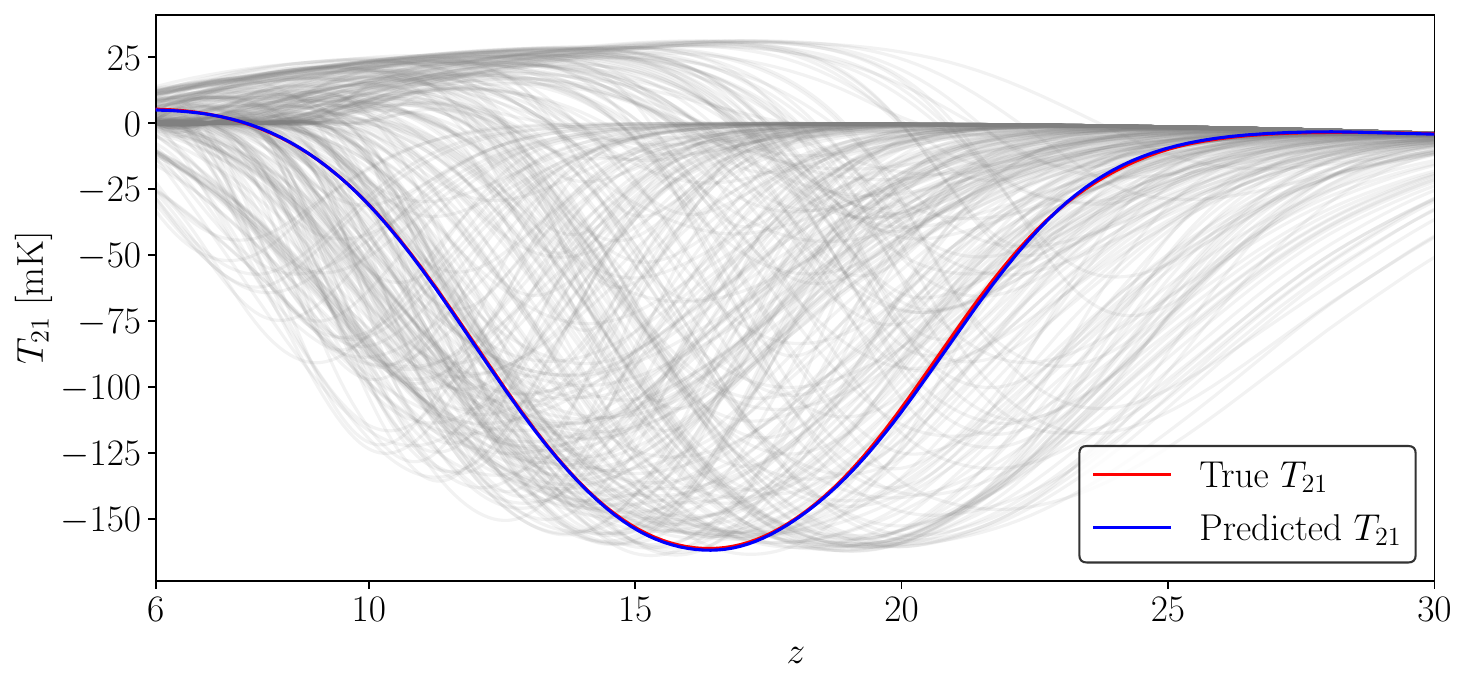}
    \end{minipage}
    \begin{minipage}{0.45\textwidth}
    \centering
    \textbf{Using the mock SKA power spectra}
         \includegraphics[scale=0.35]{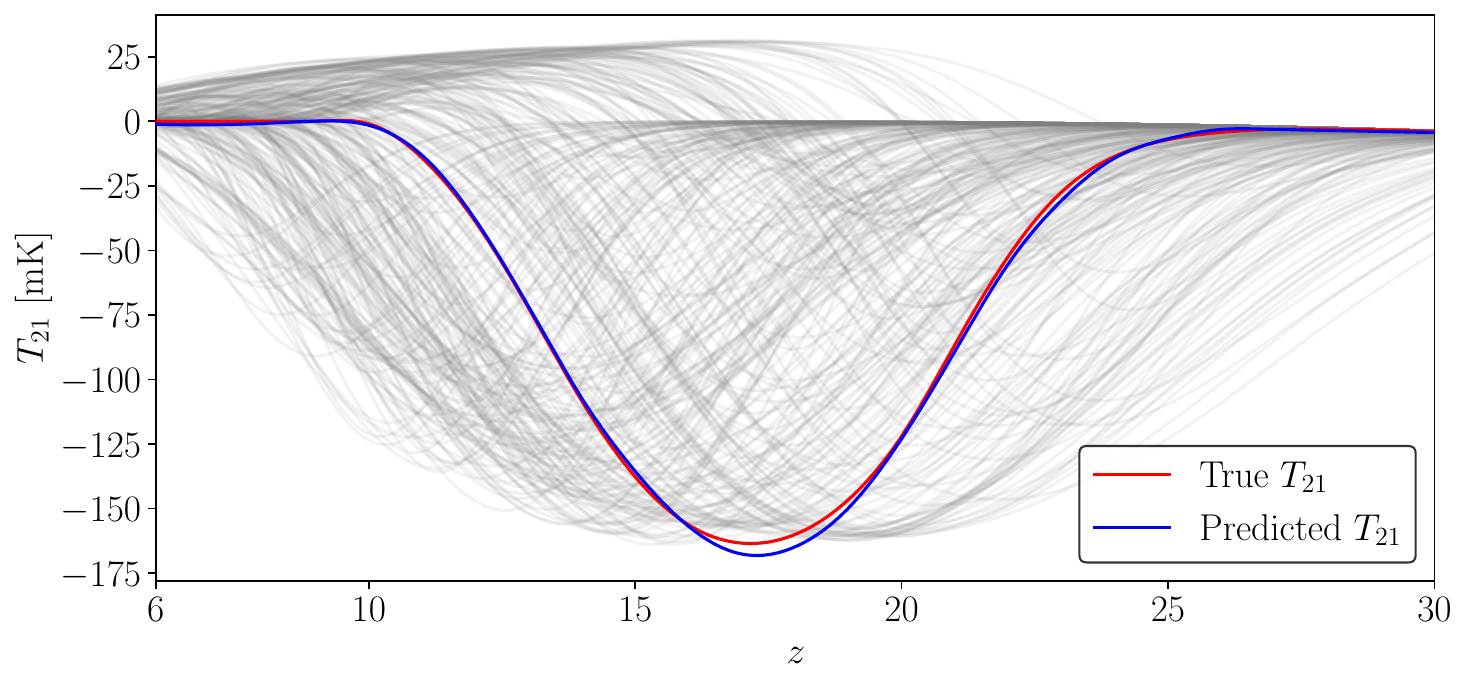}
    \end{minipage}

    \begin{minipage}{0.48\textwidth}
         \includegraphics[scale=0.35]{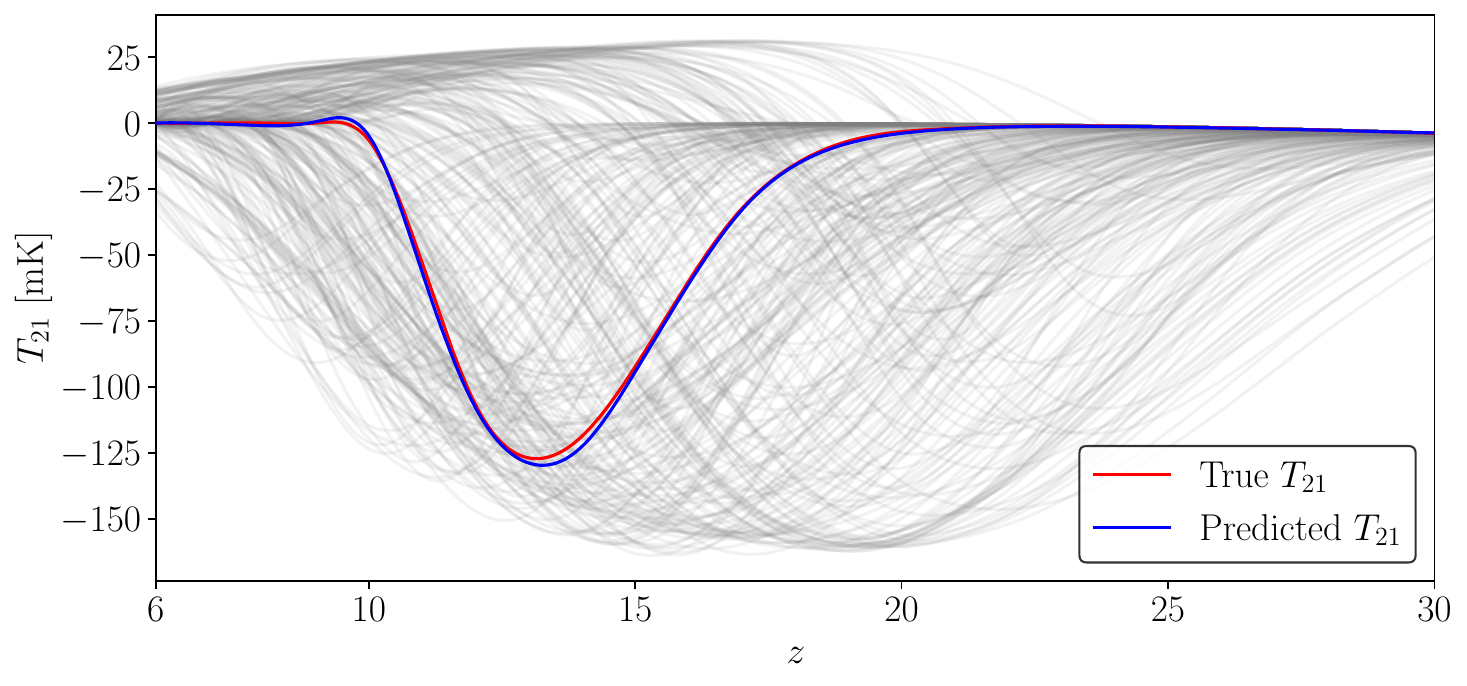}
    \end{minipage}
    \begin{minipage}{0.45\textwidth}
         \includegraphics[scale=0.35]{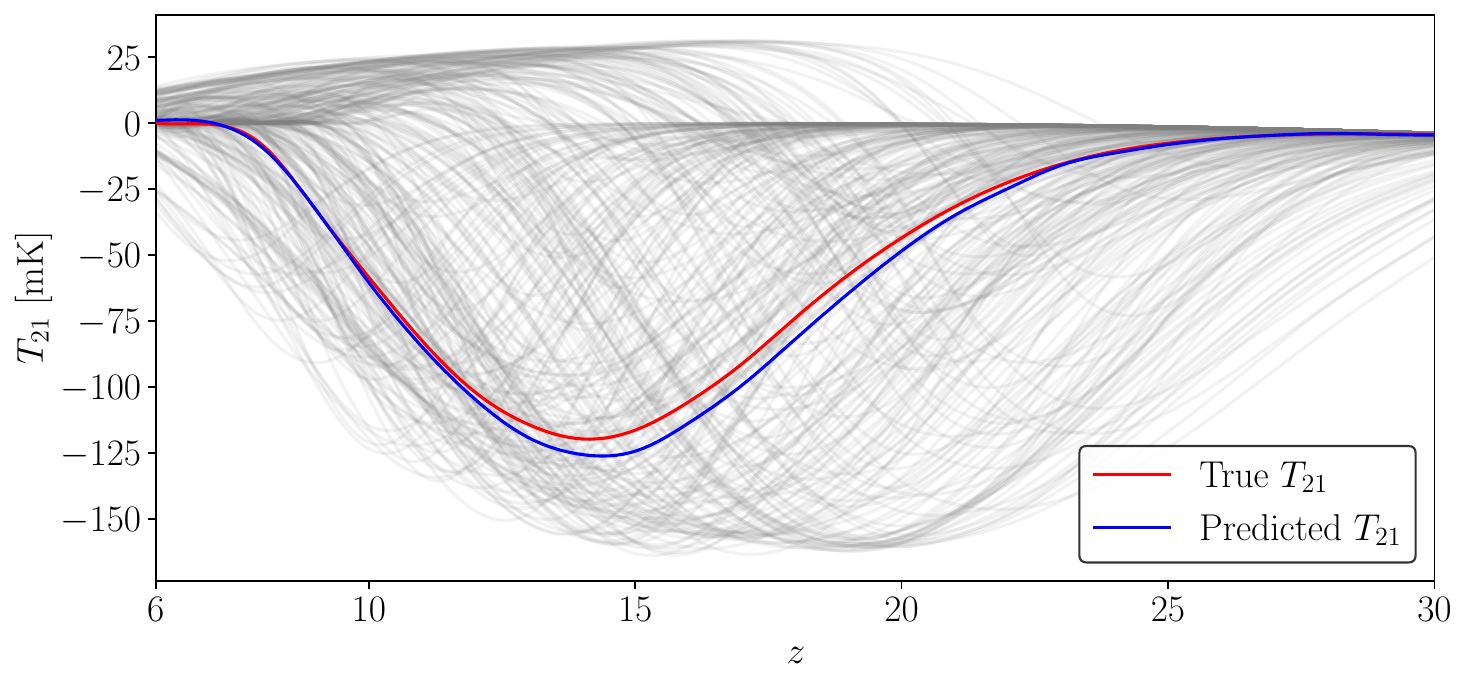}
    \end{minipage}

    \begin{minipage}{0.48\textwidth}
         \includegraphics[scale=0.35]{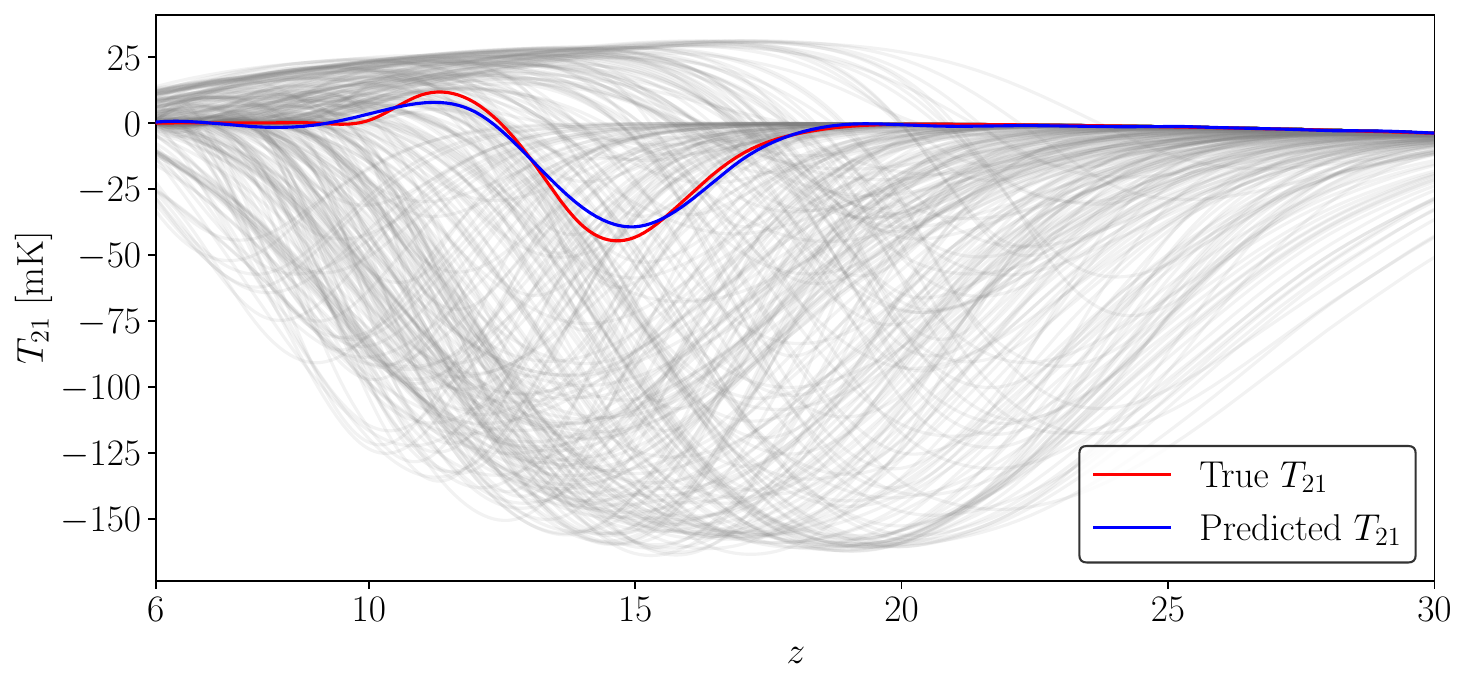}
    \end{minipage}
    \begin{minipage}{0.45\textwidth}
         \includegraphics[scale=0.35]{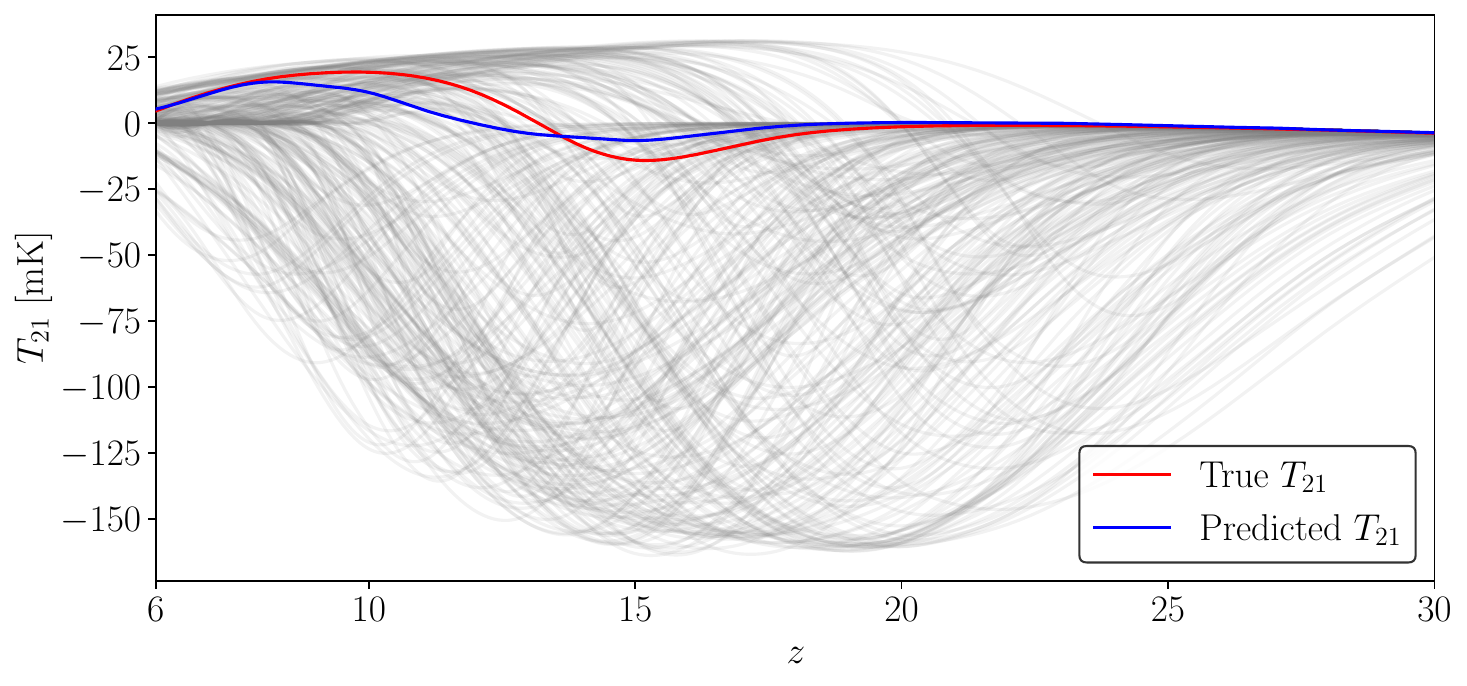}
    \end{minipage}

\caption{Comparison between the true (red curve) and predicted (blue curve) global 21 cm signal. Different models shown here illustrate particular percentiles of the error when predicting the global signal based on the power spectrum, either with or without various observational effects. The gray lines in each panel show all the samples in the test data set. \textbf{Left panels:} Here we use the clean power spectrum without any observational effects. The error illustrated by the model in each panel is as follows: Top: 10'th percentile error (0.004), middle: median error (0.010), and bottom: 90'th percentile error (0.043). The error is less than $0.061$ for $95\%$ of cases. \textbf{Right panels:} Here we predict the global signal using the mock SKA power spectrum. The error for each of the panels is as follows: Top: 10'th percentile error (0.012), middle: median error (0.035), and bottom: 90'th percentile error (0.250).}
    \label{fig:predicted_gs_clean_SKA}
\end{figure*}

For a test data set of 300 standard astrophysical models, the median value of the error is $0.0094$. Also, $95\%$ of cases have an error less than 0.061, which shows that the quality of the prediction is generally high. The predicted global signal deviates more from the actual one when the absorption trough is shallower; we note that these signals would be more challenging to detect in the first place owing to the limited telescope sensitivity. In Fig.~\ref{fig:histogram_error_gs_prediction_clean_SKA}, the dark red histogram shows the error over all the test models in the case of an ideal detection without instrumental/foreground effects. We show a few specific examples of predicting the global signal from the ANN trained using the power spectrum without any observational effects in the left panels of Fig.~\ref{fig:predicted_gs_clean_SKA}: 10'th percentile error (0.004, top panel), median error (0.010, middle panel) and 90'th percentile error (0.043, bottom panel). In each panel, the solid blue and red lines show the predicted and true global signal for the same astrophysical model parameters. It is evident from the bottom panel that even in the case of the 90'th percentile error, the location of the features was predicted well. Thus, given a noise-free power spectrum we can predict the corresponding global signal quite well, and the error in prediction is very small compared to the astrophysical uncertainty in the signal (illustrated by the scatter of gray lines).

%%%%%%%%%
%%%%%%%%%

If we use a mock SKA 21 cm power spectrum to predict the global signal using the trained network, we find that although this power spectrum includes several observational effects and is calculated over a smaller number of $z$ and $k$-bins, the trained network can still effectively predict the global signal with good accuracy. The gray histogram in Fig.~\ref{fig:histogram_error_gs_prediction_clean_SKA} shows the error over a test data set of 300 models in predicting the global signal given the mock SKA power spectrum. Here the median error is 0.035 and the error is less than 0.33 for $95\%$ of test cases. We also show a few specific examples that compare the predicted and true global signal for the same set of astrophysical parameters in the right panels of Fig.~\ref{fig:predicted_gs_clean_SKA}: 10'th percentile error (0.012, top panel), median error (0.035, middle panel) and 90'th percentile error (0.25, bottom panel). We find that the prediction is still fairly accurate, unless the absorption trough is shallow (well below 100~mK in depth).

\subsection{Prediction of the 21 cm power spectrum}\label{sec:ps_prediction}

\subsubsection{Method}

Here we attempt to do the reverse of the previous subsection. In order to map between the 21 cm power spectrum and global signal, we construct an ANN that is able to predict the 21 cm power spectrum over a wide range of redshifts and wavenumbers given the global signal over a similar range of redshifts. The network architecture comprises four hidden layers, each containing 300 neurons.
Similar data preprocessing steps have been employed, as described for the previously discussed network. After comparing the $R^2$ score and MSE for various choices of PCA components, we select 22 PCA components, since this yields the highest $R^2$ score and the lowest MSE. Consequently, the network features a 30-dimensional input layer, representing the global signal at 30 different redshifts, and a 22-dimensional output layer. By incorporating the dimensionally reduced power spectrum data set into the learning algorithm, the network effectively captures the overall signal across a broad redshift range (from 6 to 35). The original dimensions are then reconstructed using an inverse PCA transformation as a post-processing step. As a result, a reasonably good prediction accuracy is achieved. The network employs ReLU as the activation function for the hidden layers and the \textit{adam} optimizer for weight optimization.

% Similar to the network discussed previously, data preprocessing steps include transforming the power spectrum dataset to a logarithmic scale, standardizing it using the \textit{StandardScaler} class from \texttt{Scikit-learn} library, and reducing its dimensionality with PCA.
%\textit{limited-memory  Broyden–Fletcher–Goldfarb–Shanno} (BFGS)

\subsubsection{Performance analysis}

\label{sec:C}

\begin{figure*}
    \centering
    \includegraphics[scale=0.5]{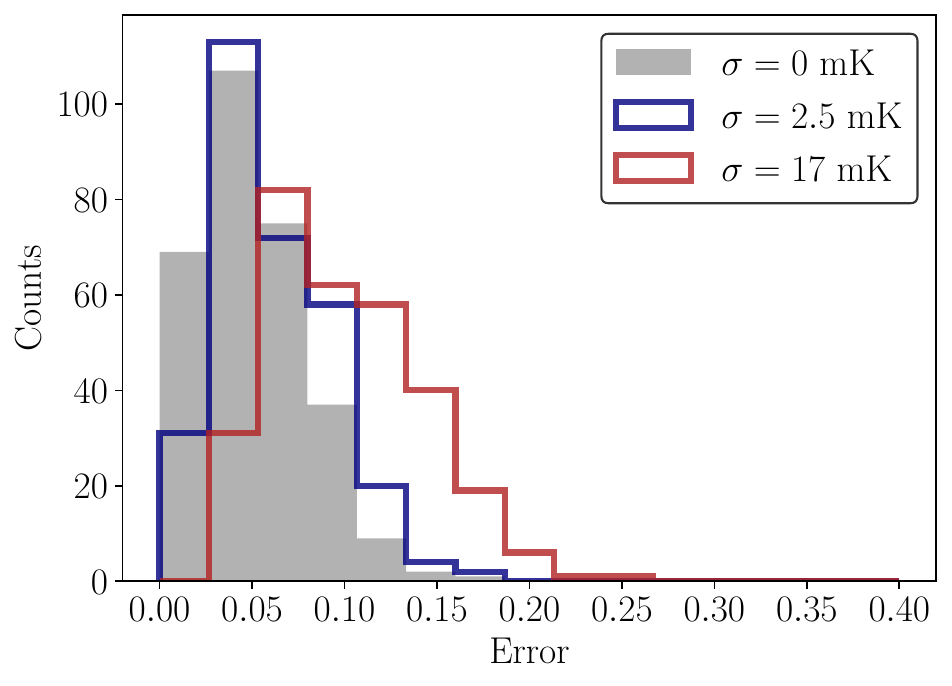}
    \includegraphics[scale=0.5]{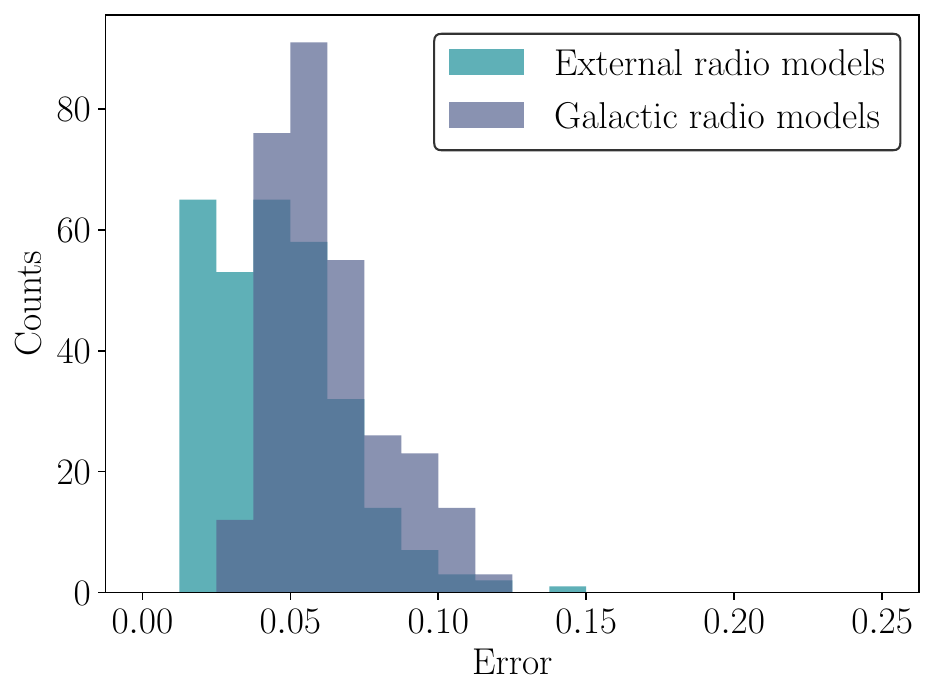}
    \caption{Histogram of the errors in the predicted 21 cm power spectrum as defined by equation~\ref{eq: ps_prediction_error}. \textbf{Left panel:} The power spectra are predicted based on the global signals using three trained networks. The results are shown for a test data set of 300 standard astrophysical models (CMB only). Here the networks are trained using global signal without any noise (gray) or with added Gaussian random noise: $\sigma=2.5$ mK, $\mu=0$ (blue) and $\sigma=17$ mK, $\mu=0$ (red). \textbf{Right panel:} The error histograms are shown in predicting the 21 cm power spectrum with an excess radio background. The teal and gray histograms show the error for the 300 models with an external radio background and 300 models with a galactic radio background, respectively. In both cases, we use the power spectrum without any observational noise.}\label{fig:histogram_error_ps_prediction_from_gs_noise_and_excess}
\end{figure*}

In order to examine the overall performance in predicting the power spectrum, we use a test data set of 300 models that were not part of the training data set. We define the error as the r.m.s.\ value of the difference between the predicted power spectrum ($\Delta_{\rm{predicted}}^2$) and the true power spectrum ($\Delta_{\rm{true}}^2$) generated from the simulation for the same parameter set, normalized by the maximum value of the true power spectrum over all $k$ and $z$:
\begin{equation}\label{eq: ps_prediction_error}
\rm{Error} = \frac{\sqrt{\rm{Mean\left[\left(\Delta^2_{\rm{predicted}} - \Delta^2_{\rm{true}}\right)^2\right]}}}{Max\left[\Delta^2_{\rm{true}}\right]} \ .
\end{equation}
Here the mean is taken over all $k$ and $z$ bins. We show the histogram of the error in gray in the left panel of Fig.~\ref{fig:histogram_error_ps_prediction_from_gs_noise_and_excess}. The median and the mean value of the error over the entire test data set are $0.045$ and $0.052$, respectively. We find that the error is less than 0.101 for $95\%$ of cases.

\begin{figure}
    \centering
    \includegraphics[scale=0.35]{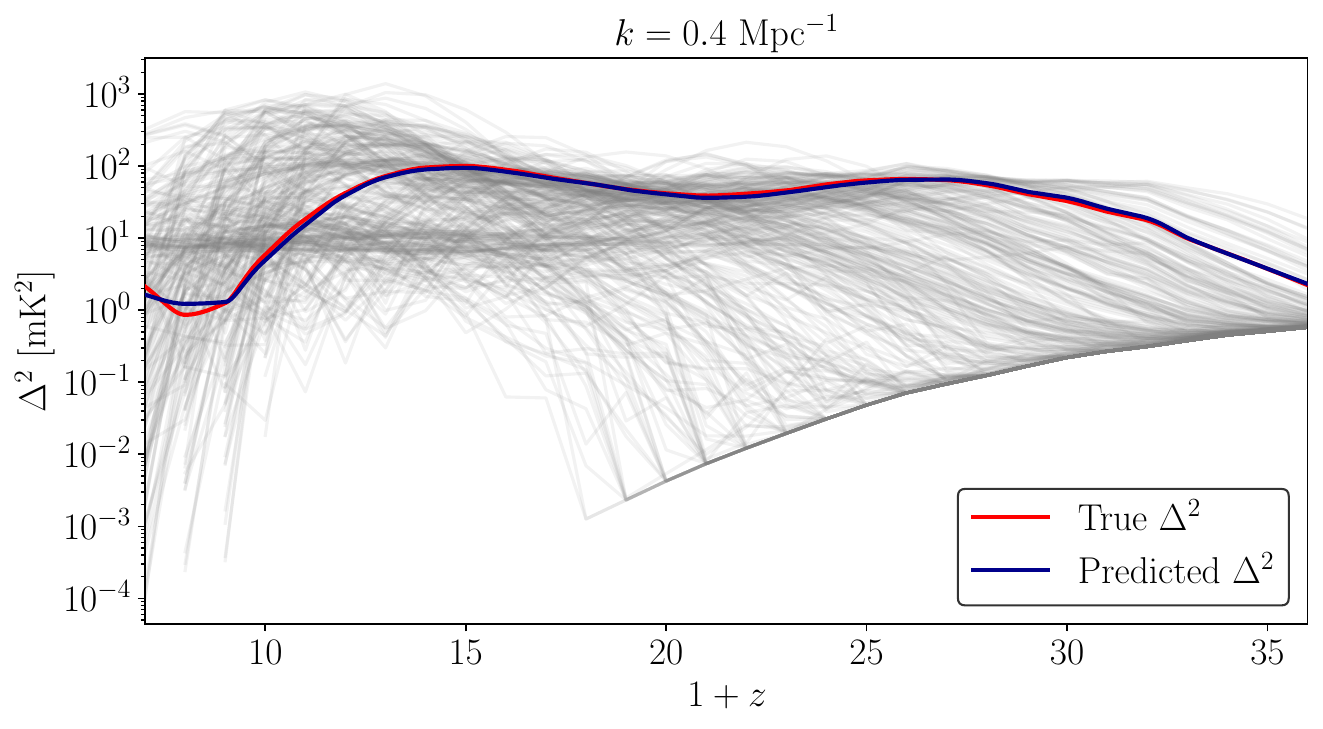}
    \includegraphics[scale=0.35]{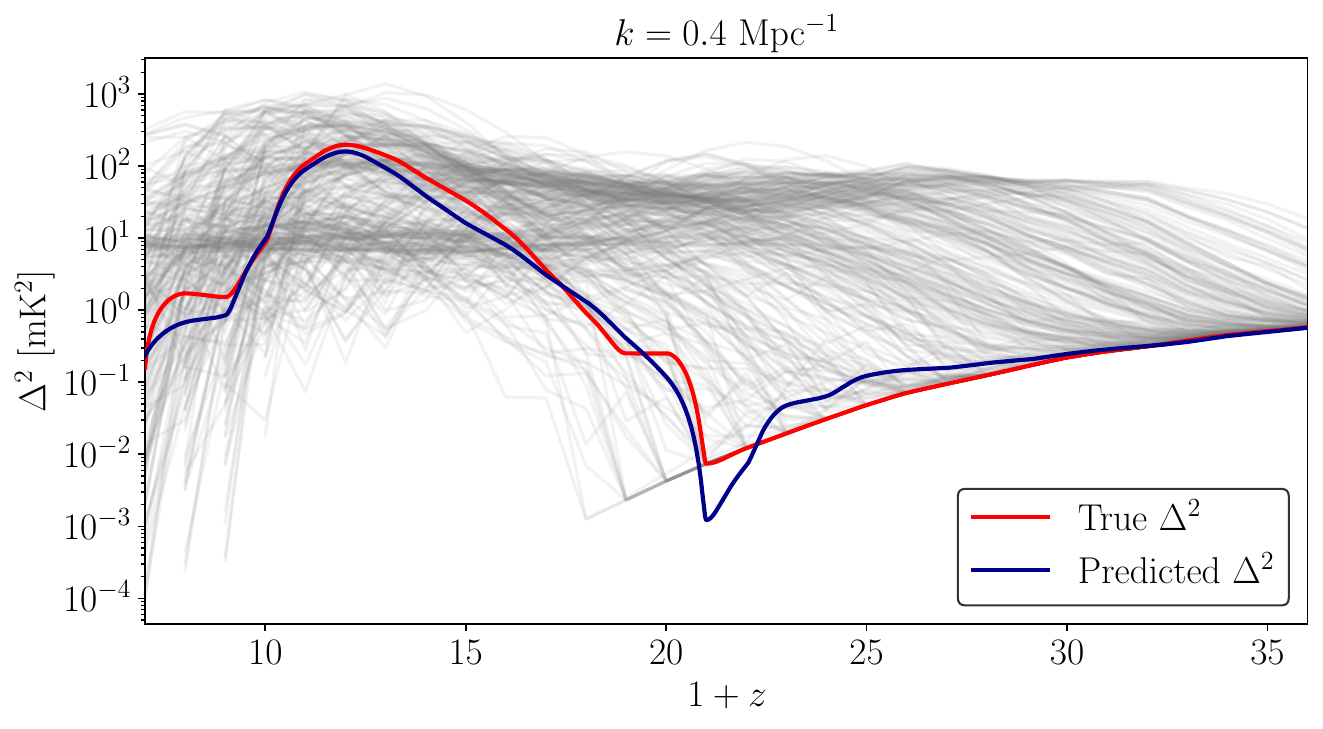}
    \includegraphics[scale=0.35]{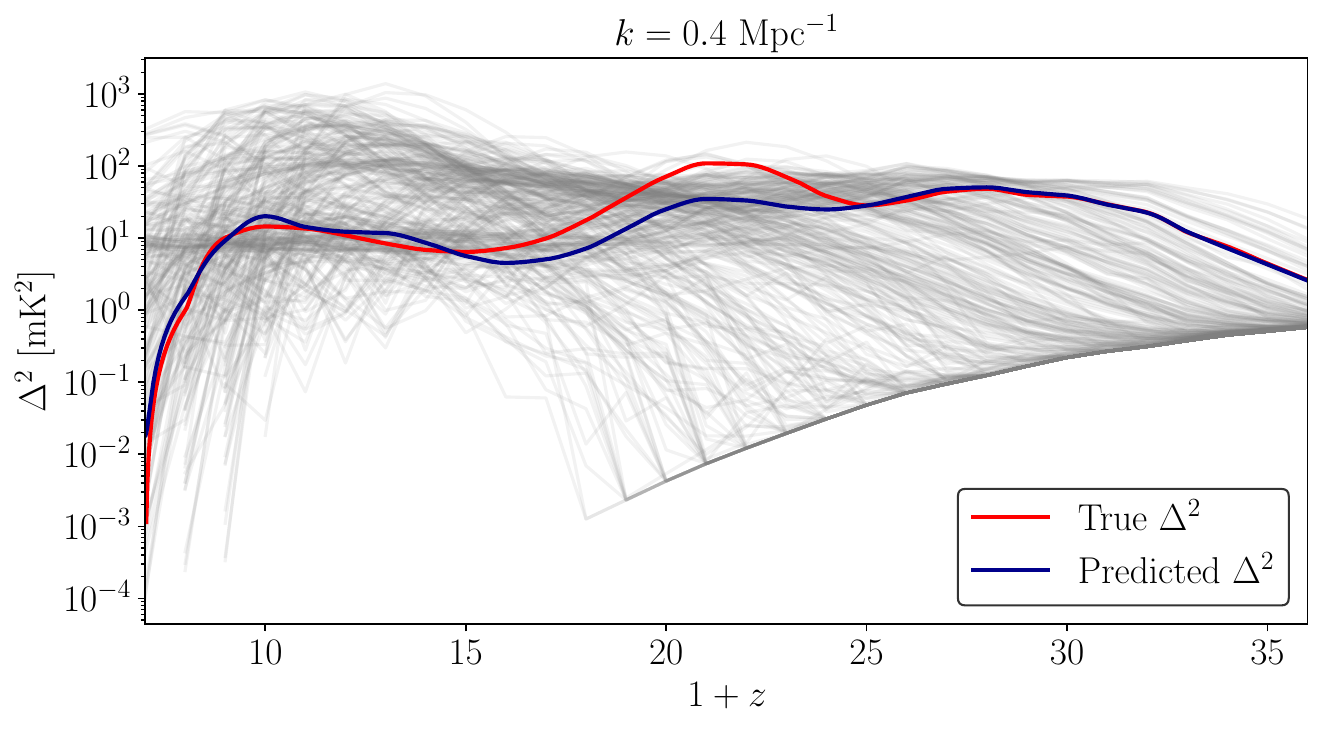}
    \caption{Comparison between the true power spectra from the simulation and the predicted power spectra from the trained network given the global signal. The level of error illustrated in each panel is as follows: Top panel: 10'th percentile error (0.020), middle panel: median error (0.045), and bottom panel: 90'th percentile error (0.092). We show the power spectra at $k = 0.4$ Mpc$^{-1}$ in all the panels. Here we use the models for the standard astrophysical scenario (CMB only). The gray lines in each panel represent all the samples in the test data set.}
    \label{fig:predicted_ps_clean}
\end{figure}

To illustrate the performance of the network in predicting the power spectrum from a given global signal, we show several specific examples in Fig.~\ref{fig:predicted_ps_clean}: 10'th percentile error (0.020, top panel), median error (0.045, middle panel), and 90'th percentile error (0.092, bottom panel). In this figure we show the power spectrum at $k = 0.4$ Mpc$^{-1}$, for illustration. We see that the prediction is generally quite good. Here we again note that the error in prediction is far smaller than the scatter due to the astrophysical uncertainty. 

In addition to predicting the power spectra within the standard astrophysical scenario, we also evaluate the overall performance in predicting the power spectra with an excess radio background over the CMB. Here we retrain the ANN separately with each class of radio backgrounds. The right panel of Fig.~\ref{fig:histogram_error_ps_prediction_from_gs_noise_and_excess} shows the histogram of errors in predicting the power spectra given the global 21 cm signals with an excess radio background. The two histograms are shown for test data sets of 300 models with an external and galactic radio background, respectively. We find that for models with an external radio excess, the error is lower than 0.081 for $95\%$ of test cases, while for models with a galactic radio background, the error is lower than 0.101 for $95\%$ of test cases, which is similar to the result for the standard astrophysical scenario. 

For illustrative purposes, comparisons between the true and predicted power spectra with an external radio background for representative cases (10th percentile error, median error, and 90th percentile error) are provided in Appendix B (Fig.~\ref{fig:predicted_ps_clean_external} and Fig.~\ref{fig:predicted_ps_clean_galactic}). In general, the prediction is reasonably good (since we are predicting the power spectrum over a wide range of wavenumbers at each redshift using only the global signal over a wide range of redshifts) and would provide a baseline for signal verification.

\subsubsection{Prediction of the 21 cm power spectrum from a noisy global signal}

\begin{table*}
\centering
\begin{tabular}{lcc} 
\hline
Global signal             & Median error & Mean error  \\ 
\hline\hline
Without any noise ($\sigma=0$)    & 0.045  & 0.052      \\
Gaussian noise with $\sigma=2.5$ mK    & 0.056  & 0.062  \\
Gaussian noise with $\sigma=17$ mK  & 0.094 & 0.100     \\
\hline
\end{tabular}
\caption{Median and mean errors in predicting power spectra given global signals without or with random Gaussian noise. These errors correspond to the histograms shown in the left panel of Fig.~\ref{fig:histogram_error_ps_prediction_from_gs_noise_and_excess}.}
\label{tab:table1}
\end{table*}

In order to understand the effect of the typical uncertainty of the global signal measurement on predicting the 21 cm power spectrum, we consider networks that are trained using the global signal with added Gaussian noise. Along with the typical sensitivity of existing global signal telescopes ($\sigma=17$ mK), we also consider an optimistic case ($\sigma = 2.5$ mK) for our analysis. In Fig.~\ref{fig:histogram_error_ps_prediction_from_gs_noise_and_excess}, we show histograms of errors (calculated using equation~\ref{eq: ps_prediction_error}) in the predicted 21 cm power spectrum, based on the global signal with added Gaussian random noise. The mean and median errors for these scenarios are listed in Table~\ref{tab:table1}. When compared to the gray histogram (for the noiseless case), it is clear that introducing Gaussian noise to the simulated global signal significantly increases the prediction error of the power spectrum. Specifically, the typical sensitivity of global signal experiments, such as the 17 mK noise level in EDGES low-band, results in a significant error, approximately 1.6 times larger than the error for an optimistic error level of 2.5 mK.

\subsubsection{Prediction of the 21 cm power spectrum using the SARAS~3 and REACH bands}

\label{sec:D}

\begin{table}
\centering
\begin{tabular}{lcc} 
\hline
Redshift bands             & Median error & Mean error  \\ 
\hline\hline
$z: 6-35$    & 0.045  & 0.052      \\
$z: 7-28$ (REACH)    & 0.050  & 0.058  \\
$z: 15-25$ (SARAS~3) & 0.064 & 0.078     \\
\hline
\end{tabular}
\caption{Median and mean errors in predicting power spectra given global signals over the REACH or SARAS~3 redshift bands. Here we consider simulated global signals without any added Gaussian random noise. These errors correspond to the histograms shown in Fig.~\ref{fig:histogram_saras3_reach}.}
\label{tab:table2}
\end{table}

\begin{figure}
    \centering
    \includegraphics[scale=0.5]{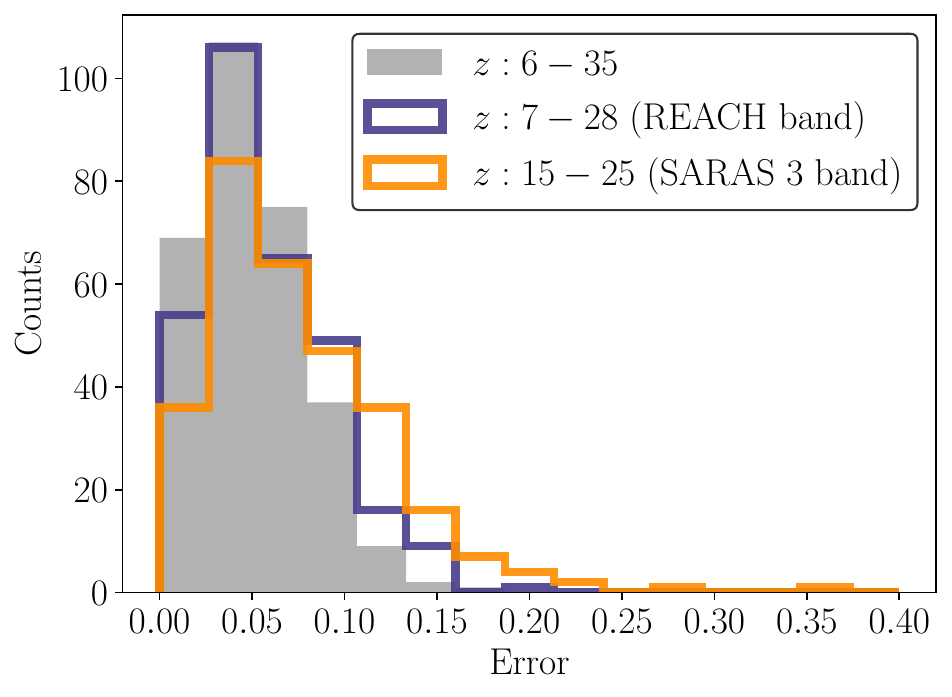}
    \caption{Same as Fig.~\ref{fig:histogram_error_ps_prediction_from_gs_noise_and_excess}, except that the power spectra are predicted based on the global signals over various redshift bands. Also, here we use the global signal samples without any added Gaussian random noise.}
    \label{fig:histogram_saras3_reach}
\end{figure}

In order to probe the effect of another typical observational constraint, we consider global signals limited to the redshift ranges of the REACH and SARAS~3 experiments. As before, we train a neural network, employ this trained network to predict the 21 cm power spectrum, and then assess the prediction performance using equation~\ref{eq: ps_prediction_error}. Here we consider the global signal without any added Gaussian noise. The analysis focuses on the same test models as discussed in Section~\ref{sec:ps_prediction}, except that the global signals is limited to specific redshift bands. Several specific examples of comparison between the true and the predicted 21 cm power spectra are presented in Appendix C, with Figs.~\ref{fig:predicted_ps_clean_reach} and \ref{fig:predicted_ps_clean_saras3} showing the results for the REACH and SARAS~3 redshift bands, respectively. To gauge the predictive capability for the power spectrum, we again quantify the errors using equation~\ref{eq: ps_prediction_error}. The histogram of the errors is displayed in Figure~\ref{fig:histogram_saras3_reach}, where the blue and dark orange histograms represent errors in the power spectrum prediction from global signals within the REACH and SARAS~3 redshift bands, respectively. The computed mean and median errors are presented in Table~\ref{tab:table2}. The error in predicting the power spectrum (both median and mean error) for the SARAS~3 band is larger by an approximate factor of 1.3 compared to that of the REACH band. Specifically, for power spectrum prediction utilizing global signals over the REACH redshift band, the median error is 0.050, while for the narrower SARAS~3 redshift band, this increases to 0.064.

\section{Conclusions}\label{sec:conclusion}

We have studied the classification of various radio backgrounds using high-redshift 21 cm data.
Typically, excess radio backgrounds are explored in the context of the EDGES low-band detection which requires extreme high-amplitude backgrounds. Here we focused on potentially more realistic cases, with moderate-amplitude radio backgrounds created by high-redshift galaxies. Such models are not usually considered but could be realized in nature. The motivation for this work was two-fold. First, we aimed to determine if an ANN classifier could pin down the nature of the radio background (here CMB-only, galaxies, or external). Secondly, we explored the possibility of utilizing an ANN to validate and verify if  detected power spectra are consistent with global signal detections, once both are detected in the future. Here are some of our main findings.

\begin{enumerate}
    \item We built two ANN classifiers that can infer the type of radio background present at high redshifts given either the power spectrum or the global signal. These networks could be very useful as part of telescope inference pipelines by helping to identify and fit the most appropriate theoretical model to the observed data. We found that the network can predict the type of radio background with an overall accuracy of $96\%$ based on a noiseless power spectrum and an overall accuracy of $90\%$ based on a noiseless global signal given as input to the trained network. We show that models with an external radio background can be distinguished with an accuracy of $100\%$ in these cases owing to the characteristic high-redshift behavior of this class of models. However, a significant fraction of galactic models are misclassified as having no radio excess. We next considered the case of signals contaminated by realistic observational effects. Using mock SKA power spectra results in a classification accuracy of $83\%$ which is still reasonably good. Adding random Gaussian noise to the simulated global signal lowers the classification accuracy. However, with an optimistic sensitivity ($\sim$ 2.5 mK) of next-generation global signal telescopes, the network can still infer the true standard astrophysical models and true external radio models  with an accuracy of $94\%$ and $93\%$, respectively. Going back to the noiseless case, we found that the accurate classification of the external radio background based on the global signal is consistent, regardless of the absorption amplitude.  For observationally motivated limits on the redshift range, in the case of the broad REACH redshift band, the classifier can distinguish the external radio background effectively, achieving an accuracy above $90\%$ for any depth of absorption deeper than 50~mK. Consequently, global signal observation through REACH-like experiments holds significant potential in discerning the presence or absence of an external radio excess in the 21 cm signal. We additionally assessed the performance of our classification model using $k$-fold cross-validation. The average classification accuracy achieved across all folds is $98.09\%$ when using the power spectrum and $96.87\%$ when using the global signal. These results highlight the robustness and reliability of our model, demonstrating its consistent performance across different validation sets. 
    \item We constructed another ANN to predict the global 21 cm signal over a wide range of redshifts given the 21 cm power spectrum. For the network where we used the power spectrum without any observational noise as the input, the error in the predicted global signal was 0.0094 (median), and less than 0.061 for $95\%$ of test cases. Given the mock SKA 21 cm power spectrum calculated over the SKA $z$ and $k$-bins, the network could still predict the global signal with a reasonably good accuracy, but with a significantly larger tail. The median error in this case was 0.035 but the $95\%$ limit was much higher (0.33).  
    \item We designed an ANN to prediction the 21 cm power spectrum over a wide range of wavenumbers and redshifts given the global signal over the same redshift range. For standard astrophysical models, we found that the prediction error was 0.045 (median) and lower than 0.101 for $95\%$ of the test models. For the models that assume a catagory of excess radio background, the errors were similar, e.g., the $95$'th percentile error was 0.082 and 0.101 for external radio models and galactic radio models, respectively. These were the cases without any noise. We explored the effect of adding Gaussian noise at various levels, or limiting the global signal information over the REACH or SARAS~3 redshift band. While the errors increased, it remained possible to predict the power spectrum with reasonable accuracy. 
\end{enumerate}

Recent studies \citep{acharya23, cyr24} indicate that soft photon heating may alter the shape and amplitude of the 21 cm signal in the presence of a strong radio background during cosmic dawn. We intend to include this effect in future work.

As it is still uncertain whether global signal experiments or radio interferometers will be able to robustly and accurately measure the 21 cm signal first, the trained network that can predict one signal from the other could play an important role of comparing the two approaches. Eventually, a demonstrated consistency between the results of the two approaches will go a long way towards convincing the broader astronomy community of the validity of the detections.

%% Please use the acknowledgment and contribution environments. This will 
%% be anonomyized when the "anonymous" style option is used. 
\begin{acknowledgments}
SS and RB acknowledge the support of the Israel Science Foundation (grants No.\  2359/20 and 1078/24). AF is grateful for the support from the Royal Society through a University Research Fellowship.

\end{acknowledgments}

\software{\texttt{Numpy} \citep{harris2020array}, \texttt{Scipy} \citep{2020SciPy-NMeth}, \texttt{matplotlib} \citep{Hunter:2007}, \texttt{Scikit-learn} \citep{scikit_learn}.}


\appendix

%\section{Selecting the optimal PCA components for the network predicting the global signal}

%In order to set an appropriate number of principal components in Section~\ref{sec:A1} we use two utility functions that measure regression performance: $R^2$ score and mean squared error (MSE). We show an example in Fig.~\ref{fig:pca_component_gs_prediction} of how these two functions vary with the number of PCA components. We select the optimum number of components where the $R^2$ score is maximized and the MSE minimized. 

%\begin{figure}
%    \centering
    %\begin{minipage}{0.4\textwidth}
%       \includegraphics[scale=0.45]{plots/r2_score_mean_squared_error_log_vs_pca_components_gs_prediction_cleanPS.pdf}

%    \caption{The impact of dimensionality reduction of the power spectrum dataset in the context of predicting the global signal given the power spectrum spectrum without any observational effects. We show how the two regression performance estimators, $R^2$ score and MSE, vary with the number of PCA components in the learning model. The dashed vertical line shows the optimum number of components.}
%    \label{fig:pca_component_gs_prediction}
%\end{figure}

%\section{Performance of the classification in the case of a noisy global signal}
\section{Impact of Stronger Noise on Global Signal Classification}

\begin{figure}
    \centering
    \includegraphics[scale=0.4]{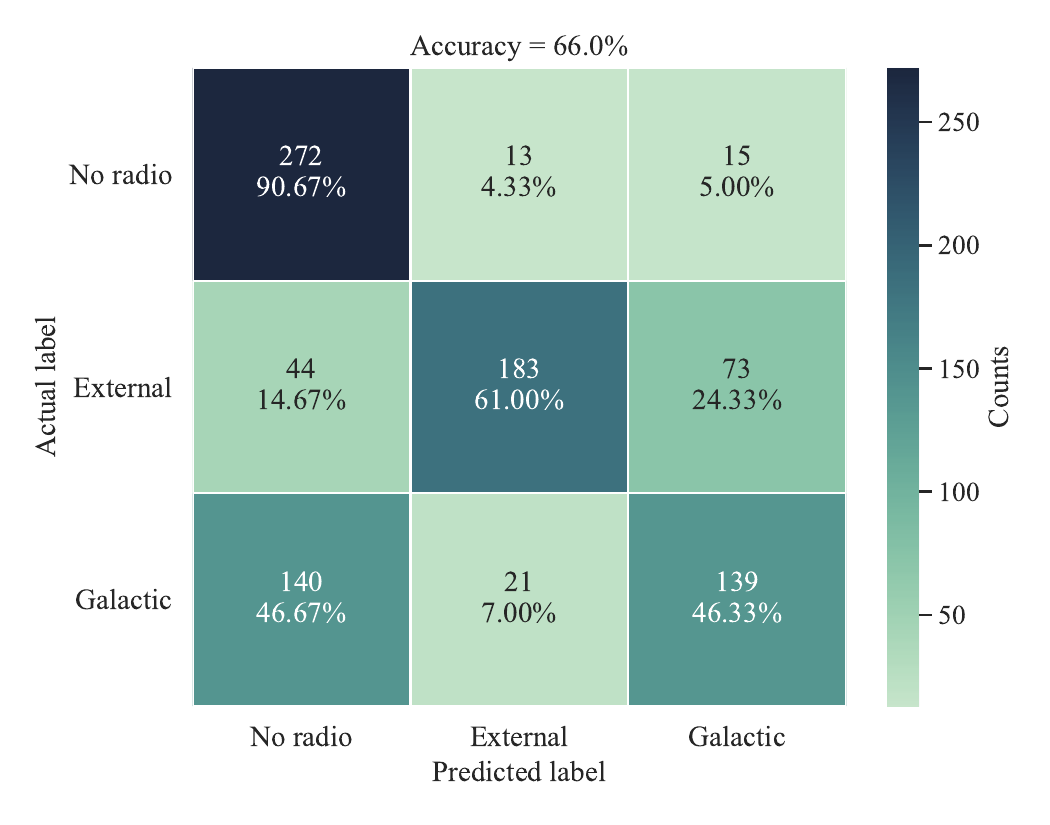}
    \includegraphics[scale=0.4]{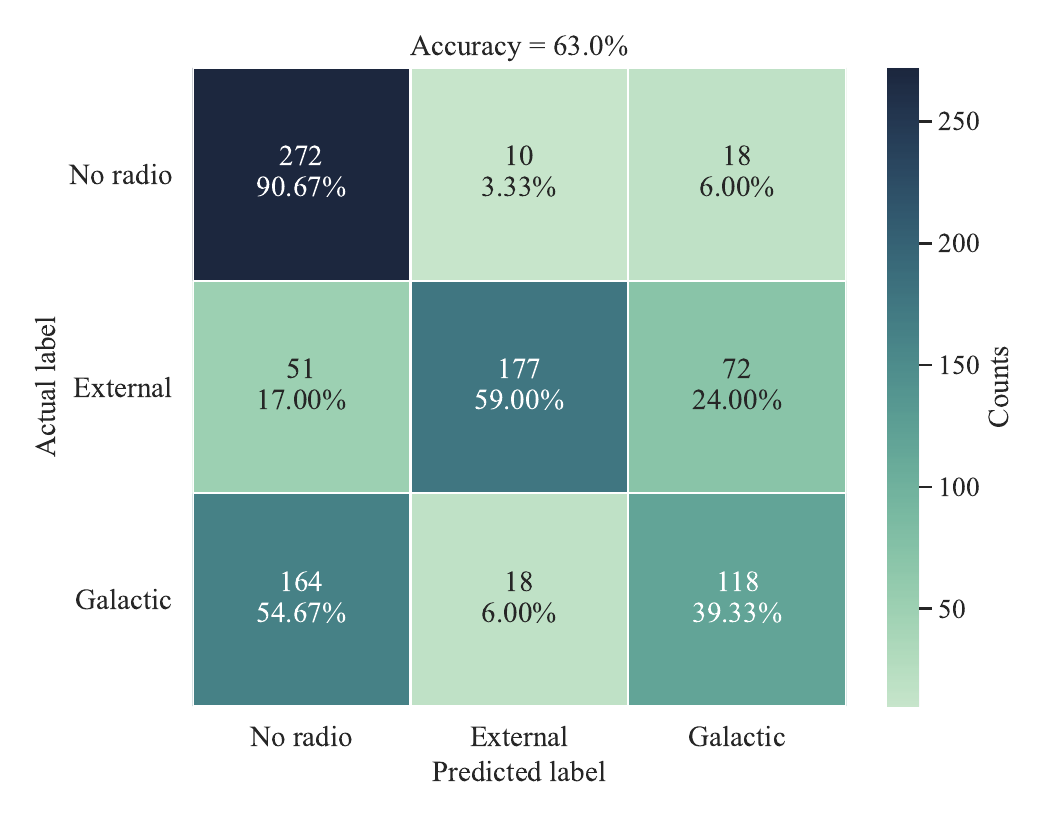}
    \caption{Classification performance procedure using the global 21 cm signal with added Gaussian noise. The left panel corresponds to $\sigma = 17$ mK and the right panel to $\sigma = 25$ mK. Stronger noise levels substantially reduce classification accuracy, particularly for the Galactic and External radio background classes.}
    \label{fig:confusion_matrix_gs_noise}
\end{figure}

In continuation of the classification analysis presented in Section~\ref{sec:classification_noisy_data}, we further examined the network’s ability to infer the type of radio background under more challenging observational conditions. Specifically, we added stronger levels of random Gaussian noise (with $\mu = 0$ and $\sigma = 17$ mK or $25$ mK) to the simulated global 21 cm signal, corresponding to noise levels representative of typical uncertainties in existing global signal experiments. As shown in Fig.~\ref{fig:confusion_matrix_gs_noise}, these higher noise cases significantly degrade the classification performance, yielding overall accuracies of $66\%$ and $63\%$, respectively. While the standard model class remains relatively robust, the Galactic and External radio background classes are much more difficult to distinguish in the presence of such strong noise.

\section{Examples of comparison between the predicted and true 21 cm power spectrum with an excess radio background}

\begin{figure}
    \centering
    %\textbf{External radio background}
    \includegraphics[width=0.32\linewidth]{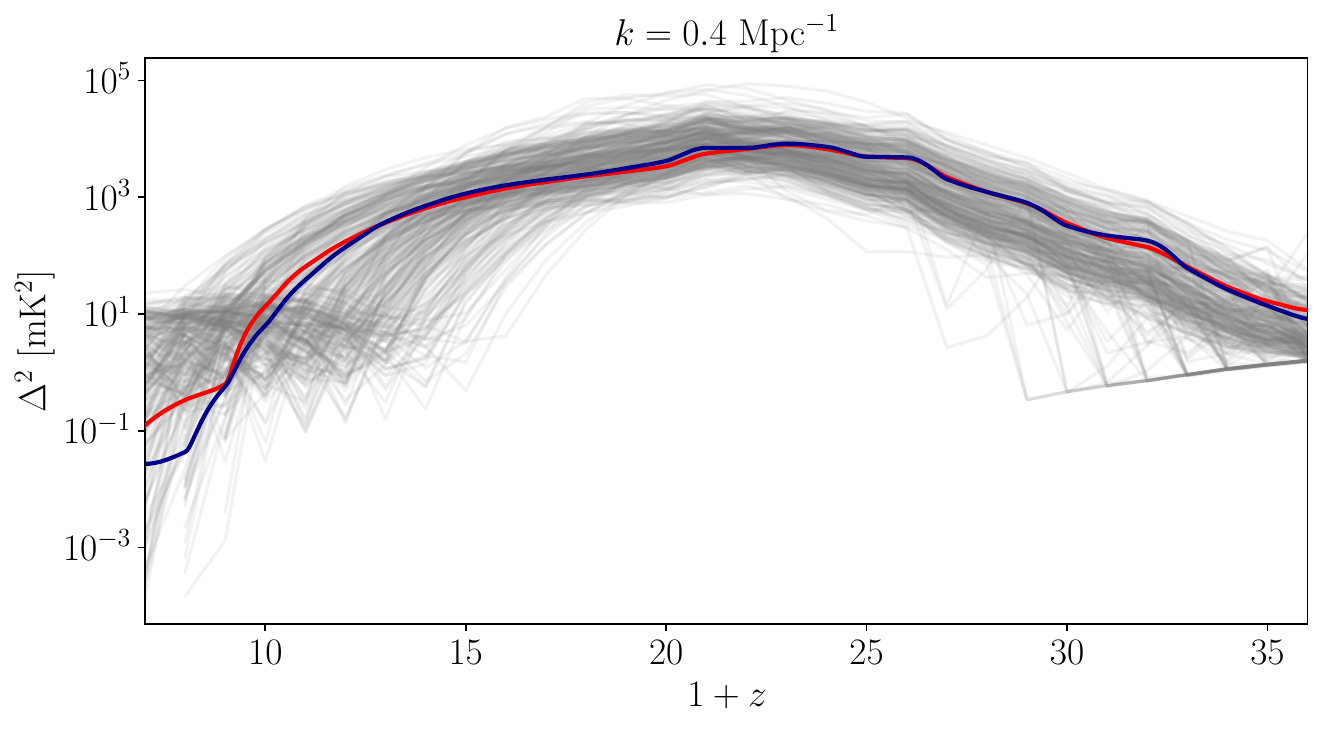}
    \includegraphics[width=0.32\linewidth]{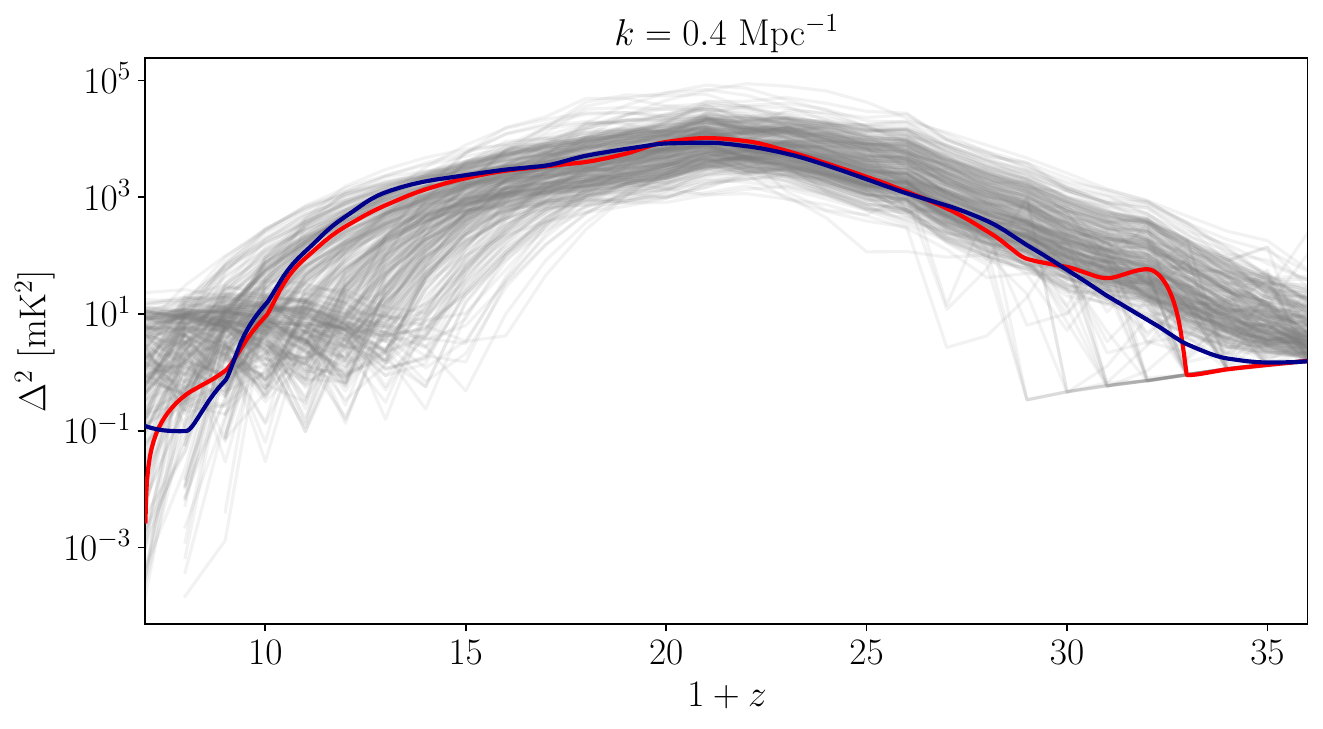}
    \includegraphics[width=0.32\linewidth]{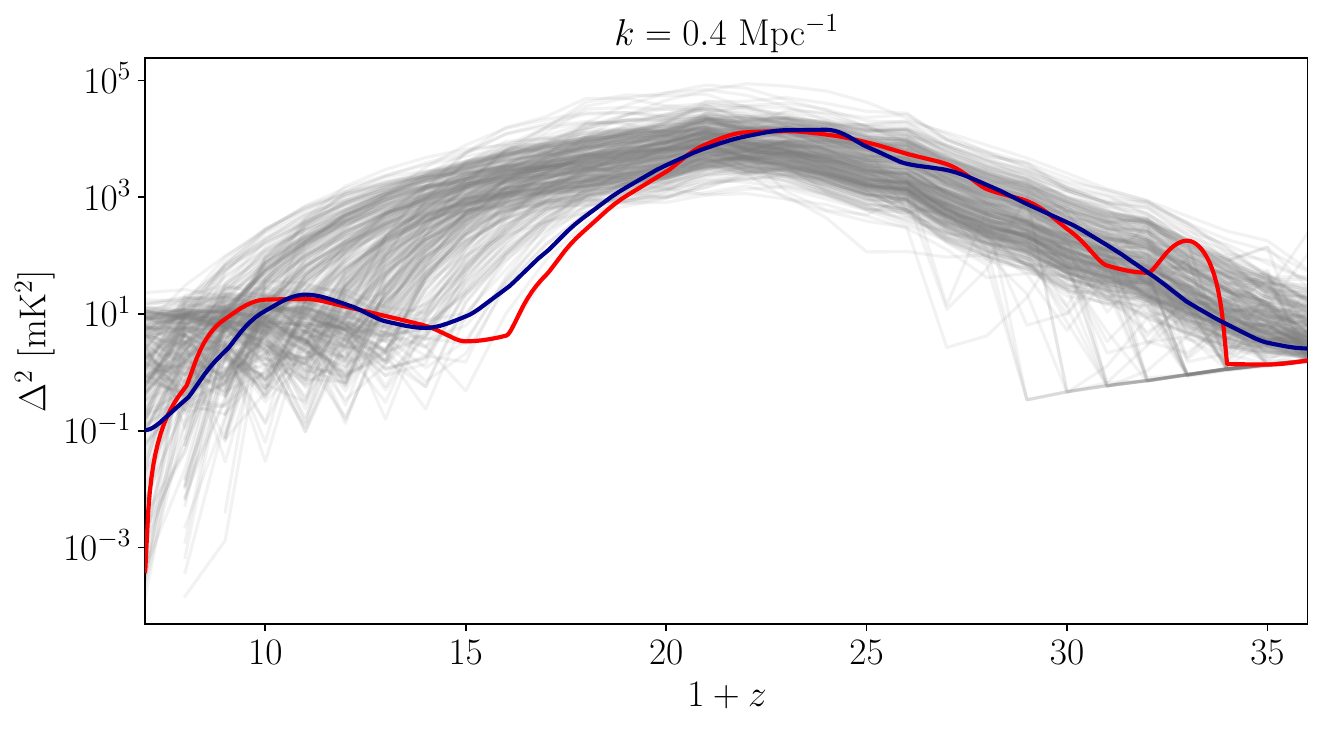}
    \caption{Comparison between the true and predicted 21 cm power spectra with an external radio background. The error level for each panel is as follows: left panel: 10'th percentile error (0.022), middle panel: median error (0.044), and right panel: 90'th percentile error (0.072). We show the power spectrum at $k = 0.4$ Mpc$^{-1}$ in all the panels. Shown here is the case without any observational noise.}
    \label{fig:predicted_ps_clean_external}
\end{figure}

\begin{figure}
    \centering
    %\textbf{Galactic radio background}
    \includegraphics[width=0.32\linewidth]{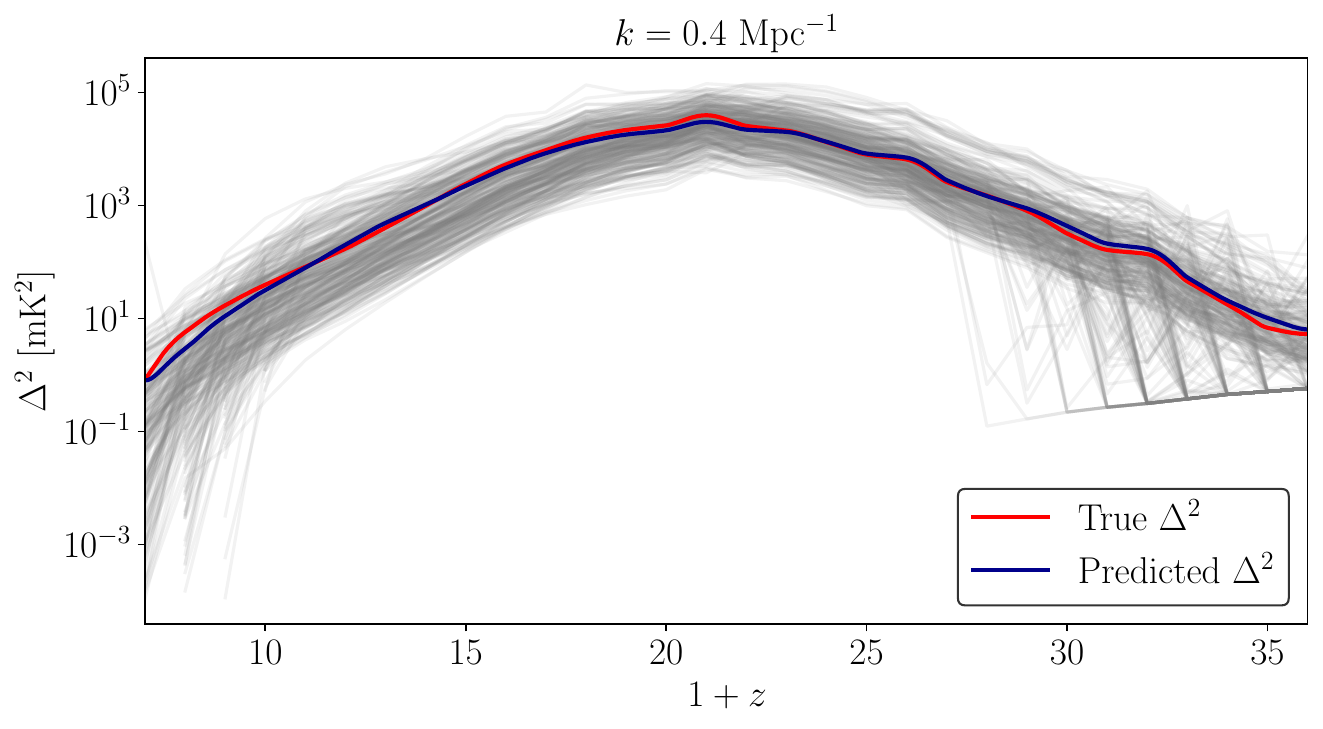}
    \includegraphics[width=0.32\linewidth]{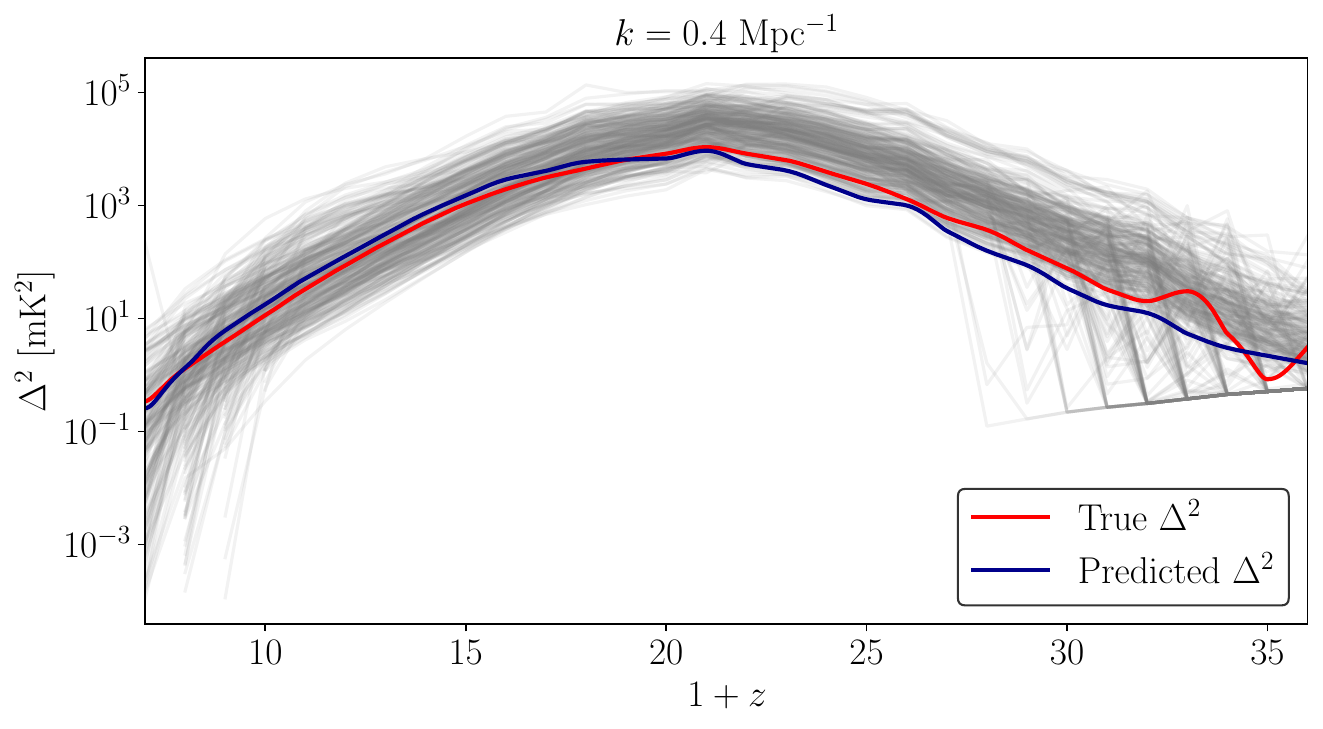}
    \includegraphics[width=0.32\linewidth]{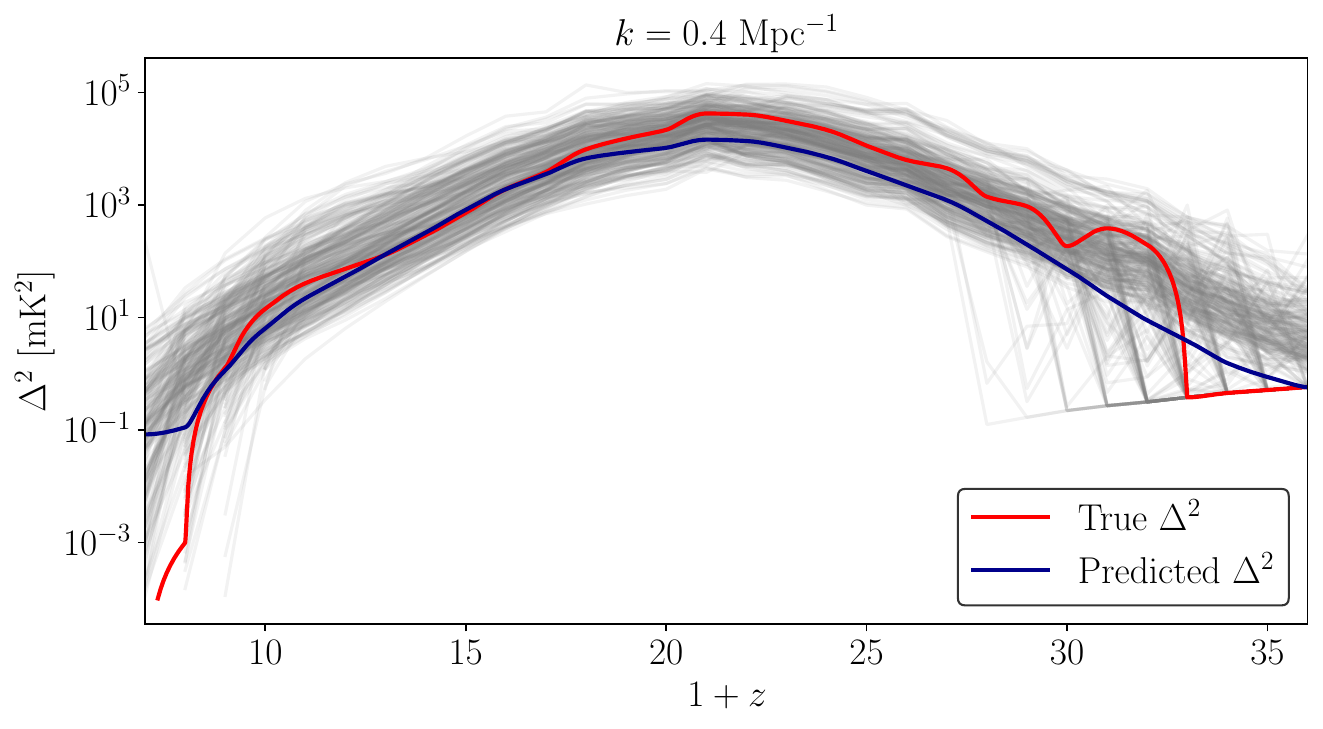}
    \caption{Comparison between the true and predicted 21 cm power spectra with a galactic radio background. The error level for each panel is as follows: left panel: 10'th percentile error (0.041), middle panel: median error (0.058), and right panel: 90'th percentile error (0.094). We show the power spectrum at $k = 0.4$ Mpc$^{-1}$ in all the panels. Shown here is the case without any observational noise.}
    \label{fig:predicted_ps_clean_galactic}
\end{figure}

In Section~\ref{sec:C}, we considered the prediction of the 21 cm power spectrum given the global 21 cm signal. To further illustrate the quality of these predictions in the case of models with an excess radio background, we present direct comparisons between the true and predicted power spectra for a few representative examples. Fig. ~\ref{fig:predicted_ps_clean_external} shows the results for models with an external radio background, while Fig. ~\ref{fig:predicted_ps_clean_galactic} presents the corresponding results for models with a Galactic radio background. These examples are based on excess radio models without any observational effects applied and are intended to highlight the network performance in different cases.

\section{Examples of comparison between the predicted and true 21 cm power spectrum from the global signal over the REACH or SARAS~3 band}

\begin{figure*}
    \centering
    %\textbf{REACH band}
    \includegraphics[width=0.32\linewidth]{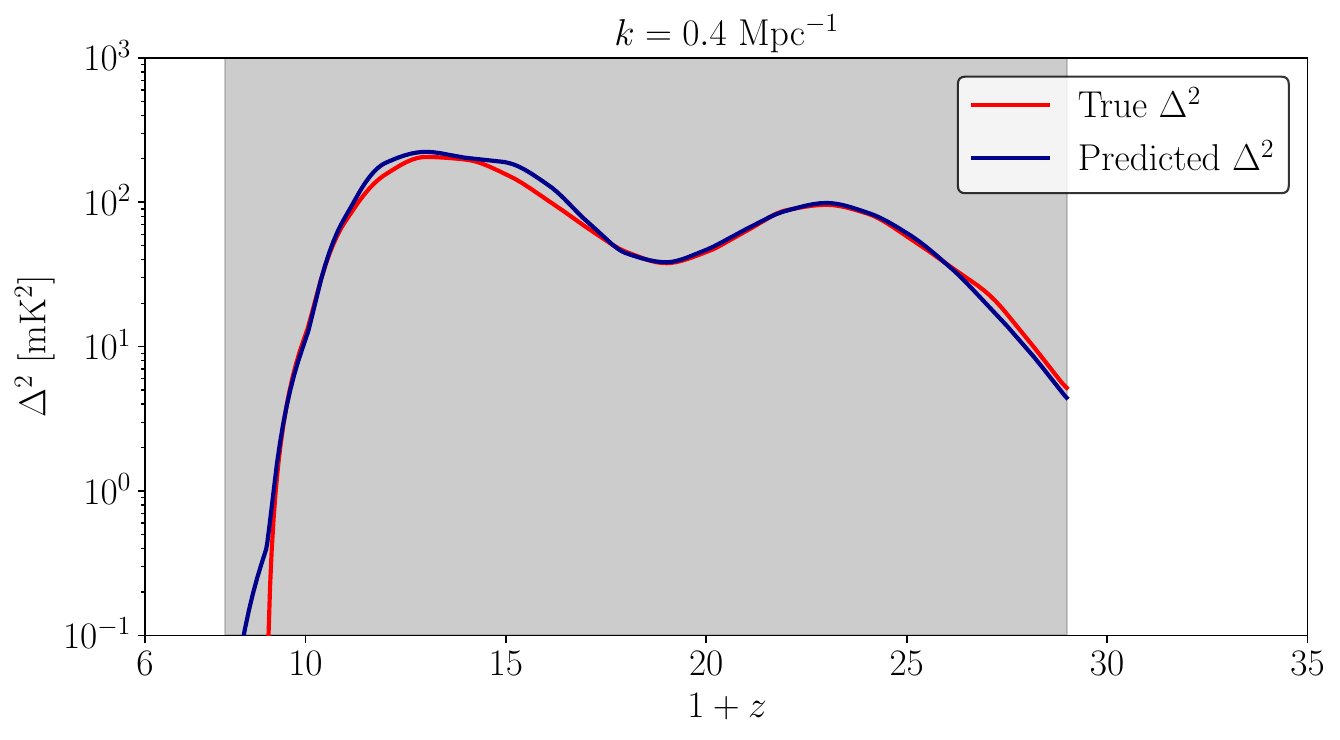}
    \includegraphics[width=0.32\linewidth]{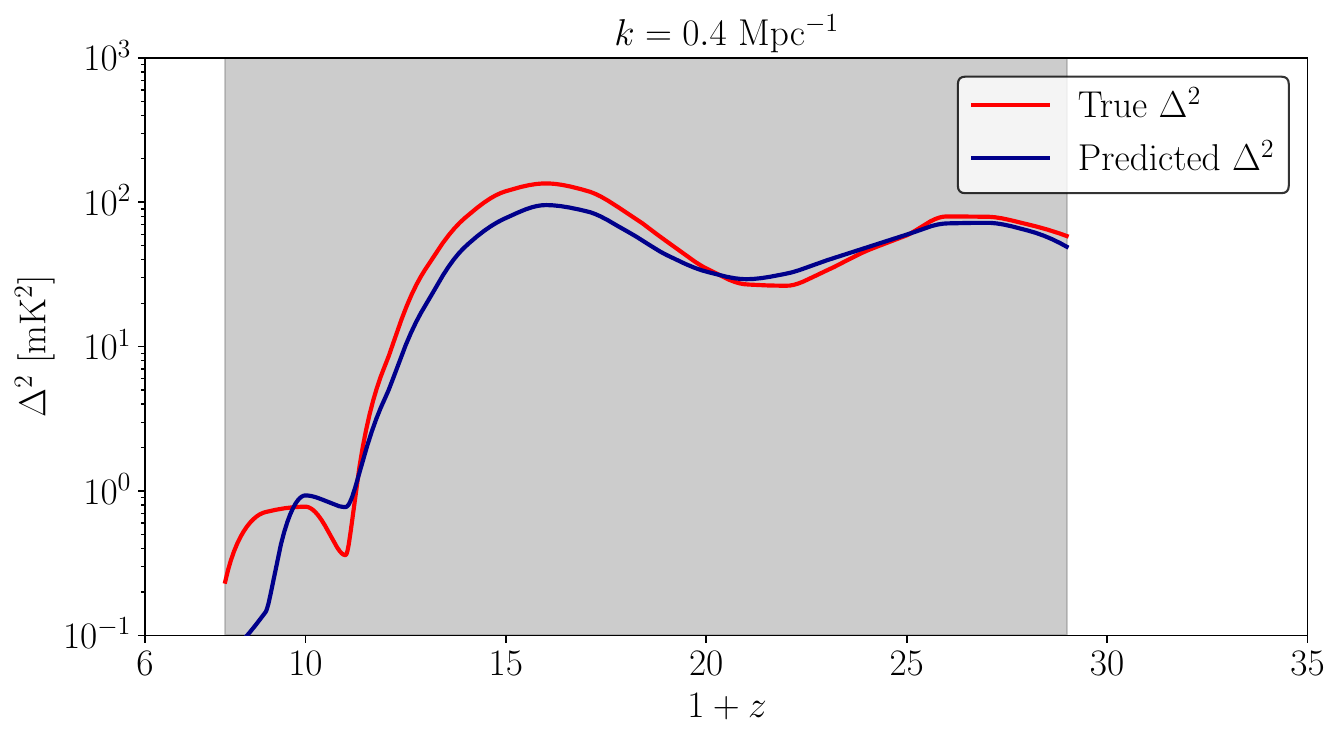}
    \includegraphics[width=0.32\linewidth]{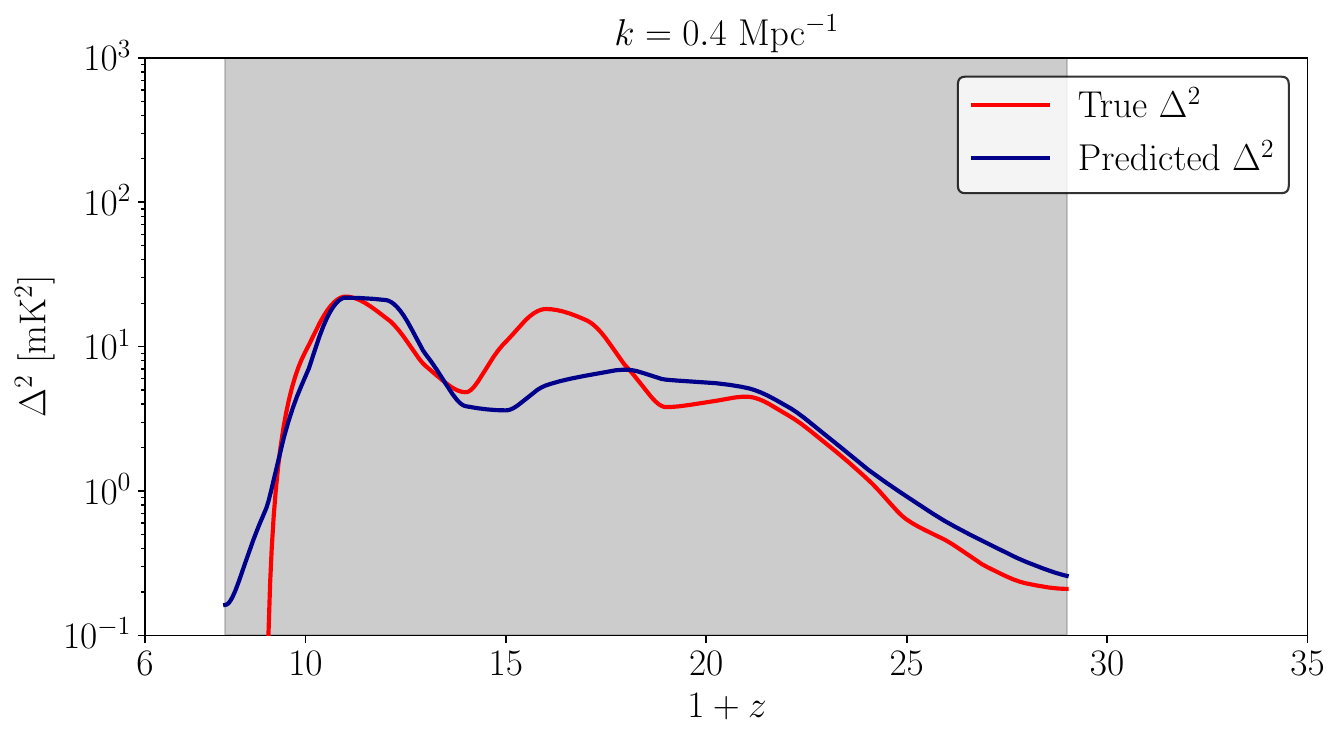}
    \caption{Comparison between the true and predicted 21 cm power spectra for standard astrophysical models using the REACH redshift band. The error level for each panel is as follows: left panel: 10'th percentile error (0.022), middle panel: median error (0.050), and right panel: 90'th percentile error (0.103). We show the power spectrum at $k = 0.4$ Mpc$^{-1}$ in all the panels. Shown here is the case without any observational noise.}
    \label{fig:predicted_ps_clean_reach}
\end{figure*}

\begin{figure}
    \centering
    %\textbf{SARAS~3 band}
    \includegraphics[width=0.32\linewidth]{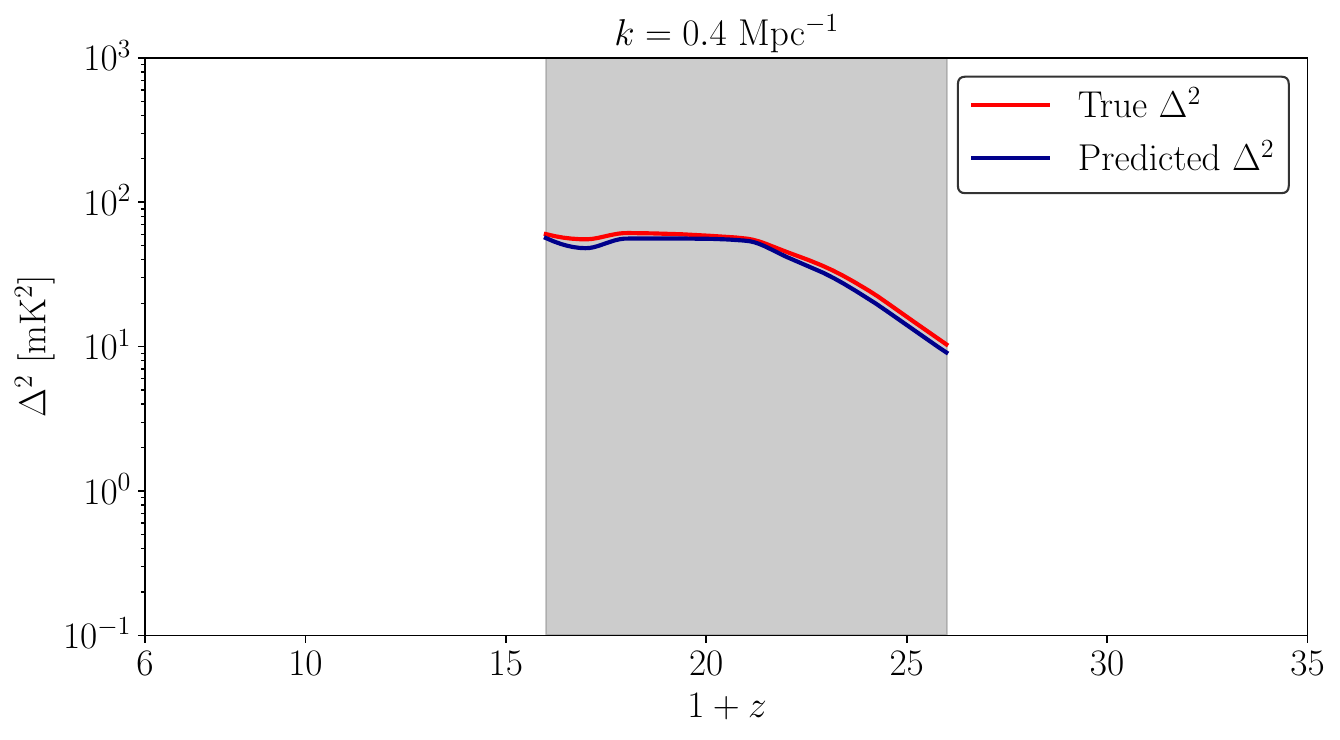}
    \includegraphics[width=0.32\linewidth]{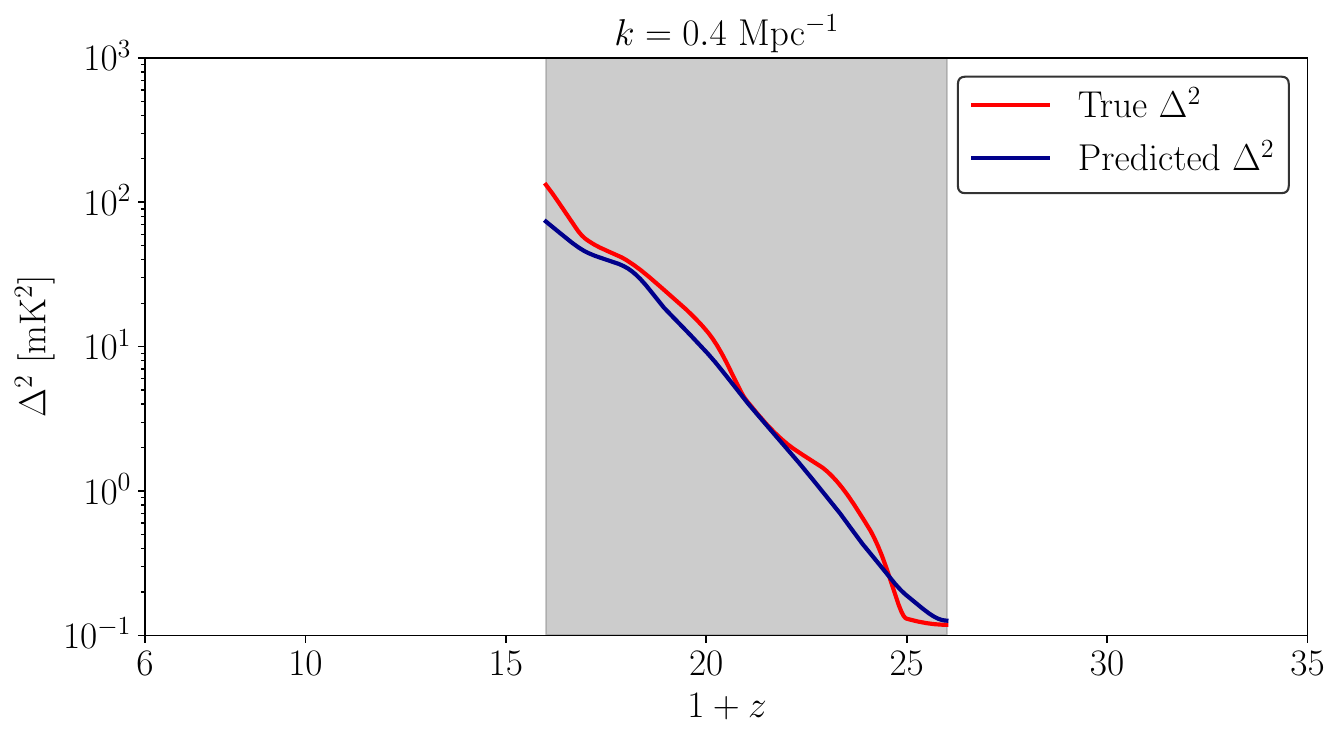}
    \includegraphics[width=0.32\linewidth]{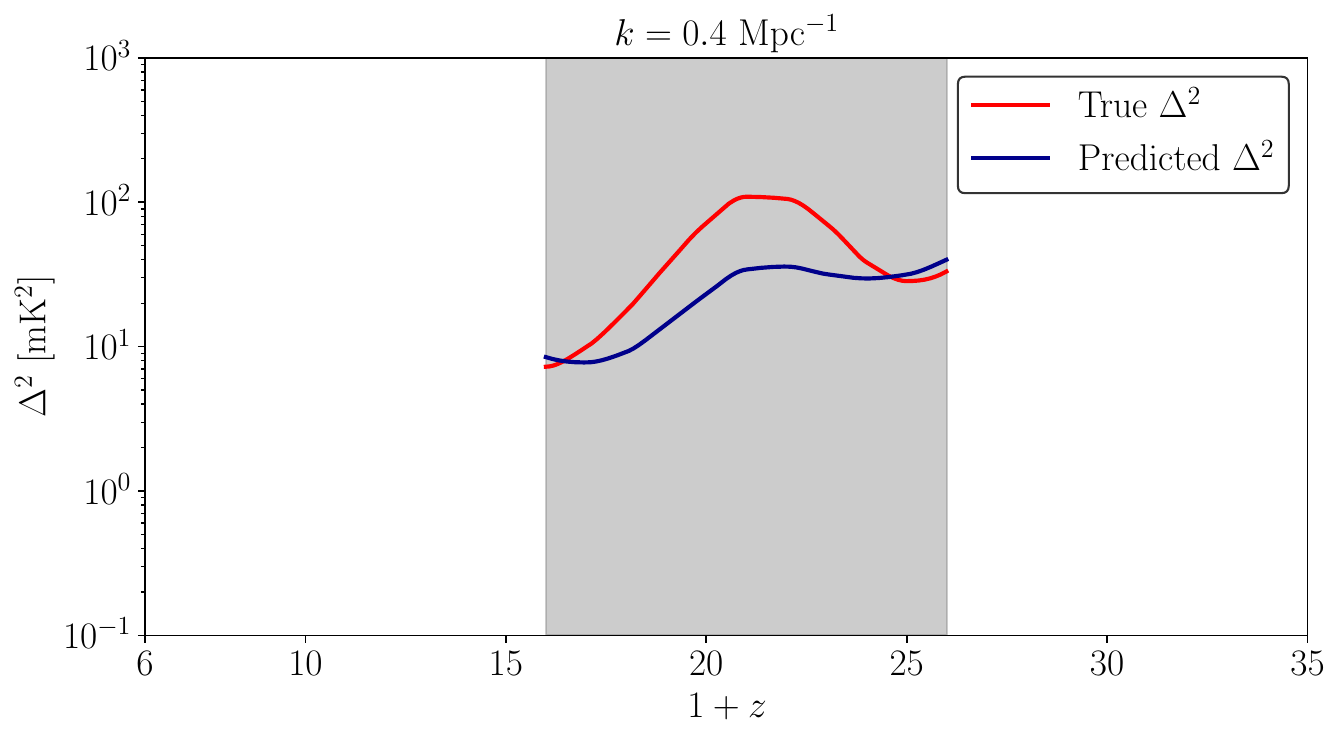}
    \caption{Comparison between the true and predicted 21 cm power spectra for standard astrophysical models using the SARAS~3 redshift band. The error level for each panel is as follows: left panel: 10'th percentile error (0.025), middle panel: median error (0.064), and right panel: 90'th percentile error (0.134). We show the power spectrum at $k = 0.4$ Mpc$^{-1}$ in all the panels. Shown here is the case without any observational noise.}
    \label{fig:predicted_ps_clean_saras3}
\end{figure}

In Section~\ref{sec:D}, we examined the prediction of the 21 cm power spectrum from the global 21 cm signal, focusing on cases where the input signal is restricted to the redshift ranges probed by the REACH and SARAS 3 experiments. Fig. ~\ref{fig:predicted_ps_clean_reach} and \ref{fig:predicted_ps_clean_saras3} present specific examples comparing the true and predicted power spectra for these two observational bands. As in the main analysis, the global signals used here do not include additional Gaussian noise.

\bibliography{sample7}{}
\bibliographystyle{aasjournal}



\end{document}